\documentclass[aps,prx,reprint,twocolumn,superscriptaddress,floatfix,nofootinbib,longbibliography]{revtex4-1}
\usepackage{epsfig,amsmath,amssymb,color,comment,physics}
\usepackage[makeroom]{cancel}
\usepackage{float}
\usepackage[normalem]{ulem}
\usepackage[english]{babel}
\usepackage{dsfont}
\usepackage{braket}
\usepackage[bookmarks=true,colorlinks,linkcolor=OrangeRed,urlcolor=NavyBlue,citecolor=RoyalBlue]{hyperref}
\usepackage[dvipsnames]{xcolor}
\usepackage{graphicx}
\usepackage{cleveref}
\graphicspath{{./Figures/}}
\usepackage{amsmath, amssymb,latexsym,amsfonts}
\usepackage{orcidlink}
\usetikzlibrary{patterns}
\usepackage{bm}
\usepackage{hyperref}
\usepackage{tabularray}
\usepackage[export]{adjustbox}

\usepackage{tikz}
\definecolor{wormred}{HTML}{C0392B}
\definecolor{loopblue}{HTML}{2C68B0}
\DeclareRobustCommand{\tailsym}{\tikz[baseline=-0.6ex]\filldraw[fill=wormred,draw=black,line width=0.4pt]
  (90:0.75ex)--(162:0.75ex)--(234:0.75ex)--(306:0.75ex)--(18:0.75ex)--cycle;}
\DeclareRobustCommand{\headsym}{\tikz[baseline=-0.6ex]\filldraw[fill=wormred,draw=black,line width=0.4pt]
  (0,0.75ex)--(0.6ex,0)--(0,-0.75ex)--(-0.6ex,0)--cycle;}
\DeclareRobustCommand{\meetsym}{\tikz[baseline=-0.6ex]\filldraw[fill=loopblue,draw=black,line width=0.4pt]
  (-0.25ex,-0.8ex)--(0.25ex,-0.8ex)--(0.25ex,-0.25ex)--(0.8ex,-0.25ex)--(0.8ex,0.25ex)--(0.25ex,0.25ex)--
  (0.25ex,0.8ex)--(-0.25ex,0.8ex)--(-0.25ex,0.25ex)--(-0.8ex,0.25ex)--(-0.8ex,-0.25ex)--(-0.25ex,-0.25ex)--cycle;}
\usepackage{multirow}

\begin{document}
\title{The 2d Ising Model on Hyperbolic Lattices and the Role of Boundary Conditions}

\author{Pallabi Dey${}^{\orcidlink{0009-0007-2460-4545}}$}
\email{pallabi.dey@saha.ac.in}
\affiliation{Theory Division, Saha Institute of Nuclear Physics, 1/AF Bidhan Nagar, Kolkata 700064, India}
\affiliation{Homi Bhabha National Institute, Training School Complex, Anushaktinagar, Mumbai 400094, India}

\author{Debasish Banerjee${}^{\orcidlink{0000-0003-0244-4337}}$}
\email{D.Banerjee@soton.ac.uk}
\affiliation{School of Physics and Astronomy, University of Southampton, University Road, SO17 1BJ, UK}

\author{Arnab Kundu${}^{\orcidlink{0000-0002-1994-3346}}$}
\email{arnab.kundu@saha.ac.in}
\affiliation{Theory Division, Saha Institute of Nuclear Physics, 1/AF Bidhan Nagar, Kolkata 700064, India}
\affiliation{Homi Bhabha National Institute, Training School Complex, Anushaktinagar, Mumbai 400094, India}

\author{Ritam Sinha${}^{\orcidlink{0000-0002-4217-9577}}$}
\email{ritamsinha.physics@gmail.com}
\affiliation{QuNu Labs Pvt.~Ltd.,~M.~G.~Road, Bengaluru 560025 Karnataka, India.}

\begin{abstract}
  We investigate the role of boundary conditions in the critical behaviour of the ferromagnetic 
Ising model on regular hyperbolic tessellations. Using worm and Metropolis simulations, we compare 
open boundaries with wired boundaries implemented through a common auxiliary spin. For open boundaries, 
the susceptibility exhibits volume-based finite-size scaling with non-mean-field scaling exponents that 
are approximately stable across four tessellations, supporting the possibility of a common universality 
class. Comparisons with Binder crossings and susceptibility peaks reveal substantial finite-size effects. 
For wired boundaries, the central-to-auxiliary spin correlator exhibits radial scaling consistent with 
a mean-field magnetization exponent. The central-spin response to a uniform magnetic field provides a further 
consistency test of $\delta=3$. Layer-resolved responses to a weak boundary field also reveal an enhancement 
of the induced magnetization towards the interior. Our results highlight the importance of boundary conditions 
and the distinction between interior and whole-system observables in characterizing critical phenomena on 
non-amenable lattices.
\end{abstract}

\date{\today}
\maketitle

\tableofcontents
\section{Introduction}

The universality of continuous phase transitions is conventionally understood in 
terms of symmetry, dimensionality and the long-distance behaviour of fluctuations
~\cite{Cardy1996, Hasenbusch_2010}. The underlying assumption here is that contributions 
from the boundary degrees of freedom become negligible in the thermodynamic limit. 
While the choice of the boundary condition can leave an imprint on the physics, it 
does not affect the standard universality associated with the continuous phase transition. 
Rigorous calculations in various lattice systems require us to consider a finite system 
in which periodic boundary conditions suppress any such finite size effects. On a 
$d$-dimensional Euclidean lattice, the volume $N$ scales as $N \sim L^d$, whereas 
the boundary scales as $N_\partial \sim L^{d-1}$, where $L$ is the linear size of 
the lattice. Therefore, in the thermodynamic limit, when $L\to \infty$, the boundary 
constitutes a vanishing fraction of the system. This plays a key role in much of 
our understanding of phase transitions 
\cite{PRIVMAN1991241, privman1990finite, PhysRevLett.28.1516, PhysRevE.72.016128, PhysRevB.74.144506}. 

 Deforming the geometry by introducing curvature may change the relative weight of 
the boundary degrees of freedom with respect to the bulk ones. Non-amenable graphs 
are such a class of lattice geometries. For example, on spaces of constant negative 
curvature, the total volume grows exponentially with geodesic radius, as do the number 
of sites on successive radial layers. Therefore, a boundary layer accounts for a finite 
fraction of the total volume even in the thermodynamic limit, unlike in flat space. 
Consequently, the boundary condition becomes integral in defining the universality 
class at the critical point. More precisely, local bulk observables and global thermodynamic 
quantities can exhibit different scaling in the infinite-volume limit. Moreover, since 
the volume no longer scales as a power of the linear size, the appropriate finite-size 
variable and the applicability of conventional finite-size scaling are not immediately 
evident. Negatively curved lattices therefore provide an apt setting for investigating 
the interplay between geometry, boundary conditions, and critical phenomena, and to 
examine the assumptions underlying conventional descriptions of universality and 
finite-size scaling.

Negative curvature spacetimes are of interest across various fields beyond statistical 
mechanics. In high energy physics, Lorentzian anti-de Sitter (AdS) spacetime forms the 
setting for the \emph{holographic correspondence}, where a gravitational theory in the 
bulk is dual to a conformal field theory living on its boundary \cite{Maldacena_1999, 
witten1998antisitterspaceholography}. In condensed matter physics, hyperbolic lattices, 
which are Euclidean AdS geometries, have been realised experimentally in circuit quantum 
electrodynamics and in electric circuits \cite{Koll_r_2019, Saa_2026, Boettcher_2020, 
PhysRevLett.128.013601}, which has motivated the development of a hyperbolic band 
theory \cite{PhysRevLett.131.226401, Sun_2024, Shankar_2024, Maciejko_2022, Boettcher_2022} 
and the study of topological phases \cite{PhysRevLett.125.053901} on hyperbolic lattices. 
Interacting systems have also been studied on hyperbolic lattices, featuring spin liquid 
phases \cite{s25y-s4fj, PhysRevLett.134.256604}. In quantum information hyperbolic lattices 
have recently been studied using tensor networks \cite{Gesteau_2022, Jahn_2022, Jahn_2020, 
samlodia2026quantumisingmodel21dimensional} and error correction codes 
\cite{taylor2023holographycellulationserrorcorrecting, Breuckmann_2016}.
Here, the extensive boundary has direct physical consequences: for superconductivity on 
hyperbolic lattices, the finite fraction of boundary sites leads to boundary-enhanced and 
even boundary-only superconducting states, with the critical temperature depending on the 
structure of the boundary \cite{bashmakov2025superconductivityhyperbolicspacesregular}. 
These developments motivate a systematic analysis of non-amenable lattices to better 
understand the interplay between the boundary conditions and the universality classes. 

 We use a controlled lattice regularization, which is described by regular $\{p,q\}$ 
tessellations of the hyperbolic plane, in which $p$-gons with $q$ polygons meet at each 
vertex. Such tessellations do not possess the translation invariance of square or cubical 
grids, but nevertheless, provide a discrete setting that encodes the effects of negative 
curvature while retaining a well-defined local connectivity \cite{Boyle_2025}. From the 
field-theory perspective, quantum field theories (QFT) have been formulated on a finite
triangulation of the hyperbolic disk, providing a lattice regularization of $\mathrm{AdS}_2$ 
in which free and interacting scalar theories reproduce the boundary correlators, as 
expected from the AdS/CFT correspondence \cite{PhysRevD.103.094507}. This construction 
was later extended to $\mathrm{AdS}_3$ \cite{PhysRevD.105.114503}. Scalar fields and 
Ising spins have also been simulated on $\{p,q\}$ tessellations, for a fixed choice 
of boundary condition, and correlation functions have been shown to exhibit a power-law 
decay on both sides of the bulk transition \cite{PhysRevD.102.034511,PhysRevD.106.054506}. 
Moreover, spin chains on the boundary of $\{p,q\}$ tilings have also been constructed 
as explicit discrete realizations of holography \cite{Basteiro_2022}. However, what remains 
relatively unexplored is the effect of the choice of boundary conditions on the phase 
transitions of the complete lattice.

 From a purely statistical-mechanics perspective, the bulk critical behaviour of the Ising 
model has been studied directly. With periodic boundary conditions (PBC) on hyperbolic 
surfaces that remove the boundary effects altogether, the Ising model falls under the 
mean-field universality class \cite{Breuckmann_2020, Krcmar_2008}. With open boundary 
conditions (OBCs), in contrast, finite-size scaling gave non-mean-field exponents, which approached 
mean-field values when the boundary layers were excluded from the observables \cite{Shima_2006, 
dey2026boundaryconditiondependentuniversality}. Several other modifications to the boundary 
can potentially give rise to distinct critical behaviour. For example, when an infinitesimal 
boundary field is applied on hyperbolic tessellations, it gives rise to an intermediate ordered 
phase that separates the paramagnetic and ferromagnetic bulk phases. 
Similarly, for the \emph{wired boundary condition} (WBC), where all the boundary
spins couple to a single auxiliary spin, the phase transition always belongs to the 
mean-field universality class \cite{f4sj-rwvj}.
Furthermore, on the so-called Cayley tree, obtained from the usual $\{p,q\}$ 
tessellations by considering $p \to \infty$, ordering in the bulk can be completely decoupled 
from that of the full system. Since closed loops are disallowed on the lattice: the interior 
orders although the free energy of the open tree remains analytic \cite{Eggarter1974}.

 Building on these studies, we compare open and wired boundary conditions within the same 
numerical framework, focusing on the resulting critical scaling and its robustness across 
different hyperbolic tessellations. In \cite{dey2026boundaryconditiondependentuniversality} 
and in this paper, we treat the boundary as a control parameter to explore how it affects 
the critical behaviour of the full lattice system.
As in \cite{dey2026boundaryconditiondependentuniversality}, we consider the 
ferromagnetic Ising model on a finite $\{p,q\}$ hyperbolic tessellation with open (OBC) and 
wired (WBC) boundary conditions, where the boundary spins are either left free or they all 
coupled to a single auxiliary spin, respectively, as we discuss in \cref{sec:model}. 
We simulate the model with the \emph{worm} algorithm for both boundary conditions and 
compare their autocorrelation times with corresponding times obtained from local Metropolis 
updates. This makes clear the necessity of using the improved worm algorithm while studying 
the critical behaviour. We then perform different scaling analyses with OBC for the 
$\{3,7\}$, $\{3,8\}$, $\{5,4\}$ and $\{6,6\}$ tessellations, examine the role of finite-size 
corrections, and correspondingly analyse the WBC transition on the different tessellations.
Our results on the intermediate \emph{Eggarter phase}, which appears when an infinitesimal 
magnetic field is applied to the boundary spins, and lies at intermediate temperatures between 
the paramagnetic and the ferromagnetic phases, is described in detail. The key feature of this 
exotic phase is that the bulk orders, while the boundary does not.

 The paper is organized as follows: in \cref{sec:geometry}, we discuss the geometry of 
hyperbolic tessellations, followed by \Cref{sec:model}, which introduces the model, the 
boundary conditions and the observables. We describe the Monte Carlo methods used to 
simulate the model with different boundary conditions, and compare their efficiency 
in \cref{sec:algorithm}. Our results for OBC, WBC and the Eggarter phase induced by a 
boundary field are described in \cref{sec:results}. We finally present our conclusions in 
\cref{sec:conclusion}.

\section{Setup}
\subsection{Geometry of hyperbolic tilings}\label{sec:geometry}
The hyperbolic plane $\mathbb{H}^2$ is the two-dimensional manifold of constant negative 
curvature $K=-1/\ell^2$, where $\ell$ denotes the curvature radius. In geodesic polar coordinates
\begin{equation}
  ds^2 = \ell^2\!\left(d\rho^2 + \sinh^2\!\rho\, d\phi^2\right), \quad \rho\in[0,\infty),\ \phi\in[0,2\pi)
\end{equation}
a circle at radial coordinate $\rho$, with geodesic radius $\ell\rho$ has circumference
$2\pi\ell\sinh\rho$ and encloses an area $2\pi\ell^{2}(\cosh\rho-1)$ \cite{PhysRevD.103.094507}.
Both grow exponentially in $\rho$ rather than polynomially, in contrast to flat space.

For the construction and visualization of the lattice we work throughout in the Poincar\'{e} disk, 
the conformal map of $\mathbb{H}^2$ onto the unit disk $\{z\in\mathbb{C}:|z|<1\}$ with metric
\begin{equation}
  ds^2 = \frac{4\ell^2}{(1-|z|^2)^2}\ dz \ d\bar z .
  \label{eq:poincare}
\end{equation}
In this metric, geodesics are circular arcs that meet the boundary $|z|=1$ orthogonally, 
and $|z|=1$ lies at an infinite geodesic distance.

A regular tessellation $\{p,q\}$ of $\mathbb{H}^2$ is a tiling by regular $p$-gons, $q$ of 
which meet at every vertex. The sign of the curvature is set by
\begin{equation}
  (p-2)(q-2)\ \begin{cases}
    <4 & \text{spherical},\\[2pt]
    =4 & \text{flat},\\[2pt]
    >4 & \text{hyperbolic}.
  \end{cases}
  \label{eq:p-qschaffli}
\end{equation}
the equality $(p-2)(q-2) = 4$ corresponds to flat lattices, for example, square $\{4,4\}$, 
triangular $\{3, 6\}$ and honeycomb $\{6,3\}$ lattices that tessellate the Euclidean flat space. 
The hyperbolic case, $(p-2)(q-2)>4$, forms an infinite family, of which we specifically focus 
on four representative hyperbolic tilings: $\{3,7\}$, $\{3,8\}$, $\{5,4\}$ and $\{6,6\}$ in 
this work. The curvature radius $\ell$ and the effective lattice spacing $a$ are related via,
\begin{equation}
    \cosh\frac{a}{2\ell} = \frac{\cos(\pi/p)}{\sin(\pi/q)} ,
    \label{eq:edge}
\end{equation}
Throughout this work, the effective lattice spacing is set to $a = 1$, so that \cref{eq:edge} 
fixes $\ell=\ell(p,q)$. A convenient measure of how strongly a tiling is curved is the 
\emph{defect angle} at a vertex,
\begin{equation}
  \delta = 2\pi - q\,\theta_p , \qquad \theta_p = \frac{(p-2)\pi}{p},
  \label{eq:defect}
\end{equation}
with $\theta_p$ being the interior angle of a flat regular $p$-gon; $\delta<0$ implies a
negative curvature, and its magnitude grows with the curvature. On the other hand, the Gaussian 
curvature $K_{\rm Gauss}$ is given by\cite{PhysRevLett.125.053901},
\begin{align}
    K_{\rm Gauss} = \frac{\delta}{A_{\rm poly}} = -\frac{\pi p }{A_{\rm poly}}\left(1-\frac{2}{p}-\frac{2}{q}\right)~~,
\end{align}
where $A_{\rm poly}$ is the area of each $p$-sided unit polygon in the lattice. Given 
that the thermodynamic limit of the hyperbolic lattice is independent of $A_{\rm poly}$, 
one can scale away the area by assuming 
$K_{\rm scaled} = K_{\rm Gauss} ~A_{\rm poly}/p$.
The scaled Gaussian curvature 
$K_{\rm scaled}$ is therefore given only in terms of the tiling parameters $\{p,q\}$ \cite{f4sj-rwvj}:
\begin{align}
    K_{\rm scaled} = -\pi\left(1-\frac{2}{p}-\frac{2}{q}\right)~~.
\end{align}
The curvature $K_{\rm scaled}$ is invariant under $p \leftrightarrow q$, i.e., 
$K_{\rm scaled}(p,q) = K_{\rm scaled}(q,p)$. Among the four tilings we have studied here, 
$|K_{\rm scaled}|$ is smallest for $\{3,7\}$ tessellation and largest for $\{6,6\}$ tessellation, 
with $\{3,8\}$ and $\{5,4\}$ tessellations lying in between.
\begin{figure}[!tbh]
    \centering
    \includegraphics[width=\linewidth]{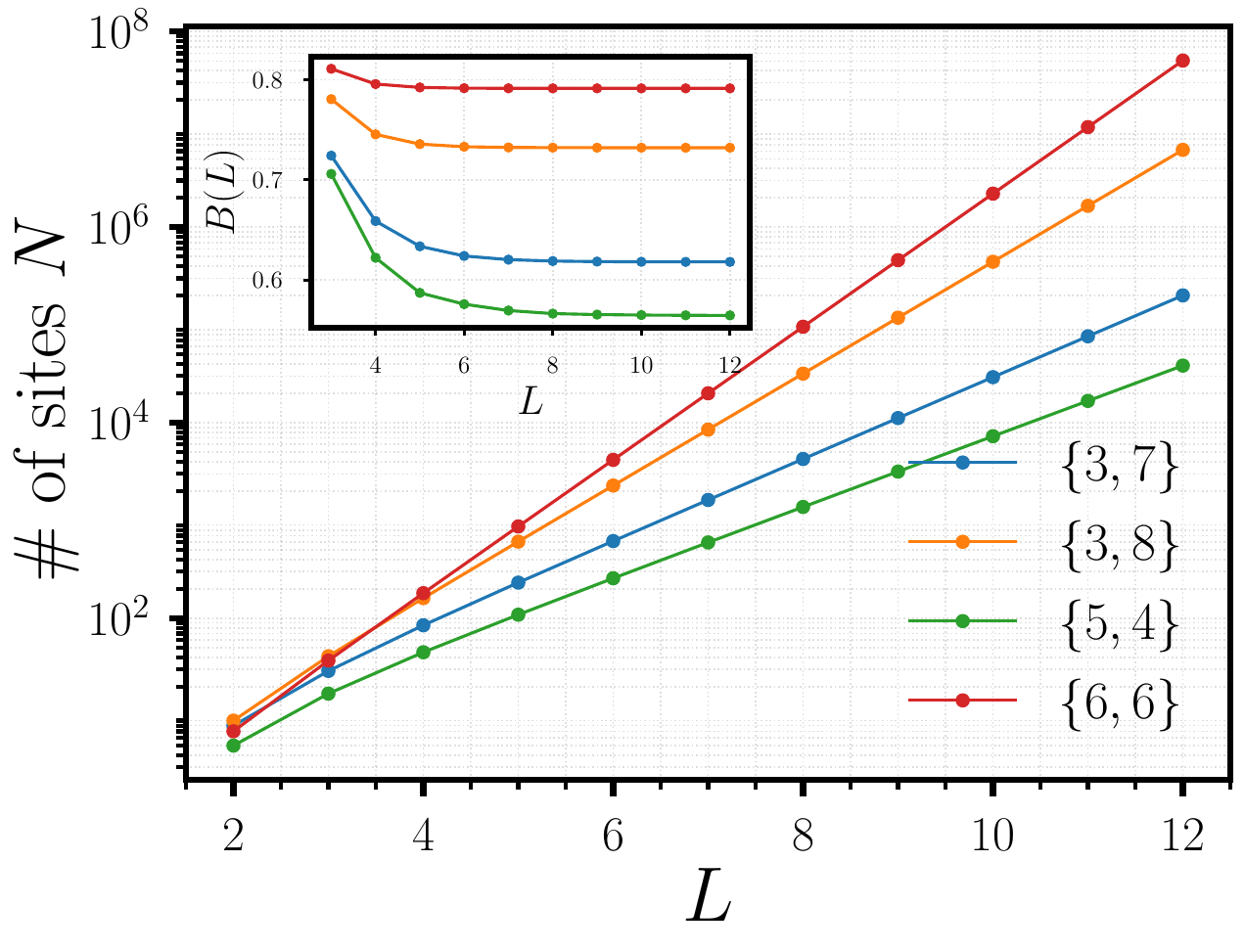}
    \caption{Exponential growth of the number of sites $N$ with the layer $L$ for the 
    $\{3,7\}$, $\{3,8\}$, $\{5,4\}$, and $\{6,6\}$ tilings. The linear behavior on a 
    semi-logarithmic scale confirms $N(L)\sim e^{c_{p,q}L}$, where the $\{6,6\}$ tessellation, 
    with the largest $|K_{\rm scaled}|$, shows the steepest slope. Inset: The boundary 
    fraction $B(L)$ approaches a nonzero constant (tiling dependent) at large $L$.}
    \label{fig:N-L}
\end{figure}

We use the HYPERTILING package \cite{SciPostPhysCodeb.34, SciPostPhysCodeb.34-r1.3} to construct 
finite lattices on regular hyperbolic tessellations. Starting from a central polygon (the first layer), 
we iteratively add adjacent polygons to form successive concentric layers. The construction is 
truncated at layer ($L_{\rm{max}}$), which determines the size of the resulting finite lattice.
The number of sites grows exponentially with each layer as $N(L)\sim e^{c_{p,q} L}$, at a rate 
$c_{p,q}$ defined by the curvature \cref{fig:N-L}. Since the fraction of sites on the boundary 
relative to the total sites approaches a non-zero, curvature dependent value as the volume increases, 
\begin{equation}
  B(L)=\frac{N(L)-N(L-1)}{N(L)}\;\xrightarrow{L\to\infty}\; 1-e^{-c_{p,q}} = 1 - \frac{1}{\lambda_{p,q}},
  \label{eq:bfrac}
\end{equation}
as shown in the inset of \cref{fig:N-L}. The growth factor $\lambda_{p,q}>1$, which determines the 
exponential growth rate through $c_{p,q}=\log\lambda_{p,q}$, is the largest real root of the 
denominator polynomial of the vertex growth series for the $\{p,q\}$ tessellation, as discussed 
in \cite{Barthold2002}. Thus, the boundary $B(L)$ is not negligible in the 
thermodynamic limit, unlike a flat lattice. The boundary therefore cannot be treated as a 
finite-size artifact: it persists in the thermodynamic limit, and the boundary condition becomes 
a physical input that can control the critical behavior. For the rest of the paper, we will
use the notation $ \partial V = N(L) - N(L-1)$, which counts the number of sites in the last layer.

\subsection{Model and boundaries }\label{sec:model}
\begin{figure}[!tbh]
    \centering
    \includegraphics[width=\linewidth]{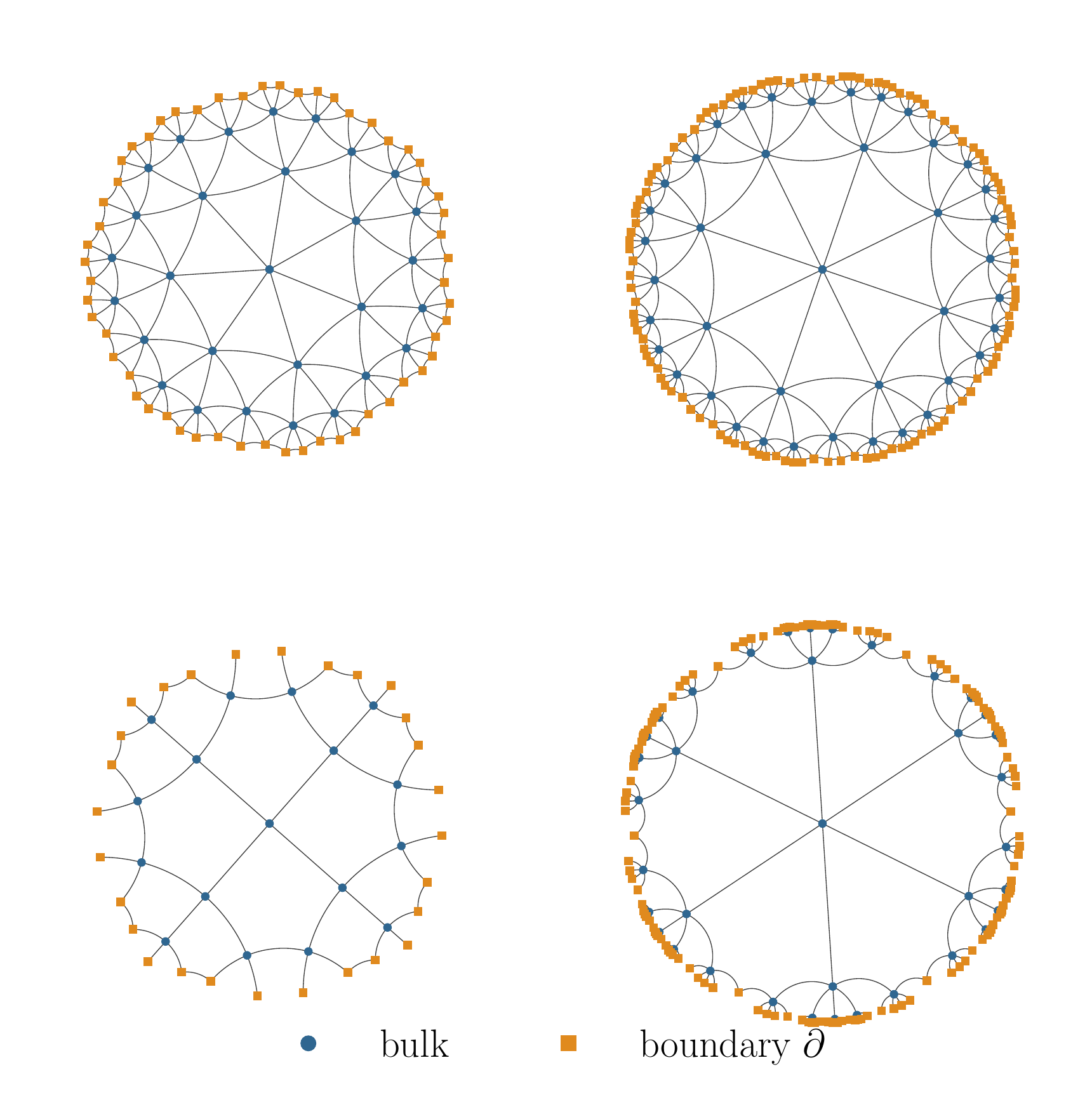}
    \caption{Regular hyperbolic tilings in the Poincar\'e disk with open boundaries, for the  $\{3,7\}$ 
    (top-left), $\{3,8\}$ (top-right), $\{5,4\}$ (bottom left) and $\{6,6\}$ (bottom right) tessellations.}
    \label{fig:tiling}
\end{figure}

We consider the conventional ferromagnetic Ising model on these tessellations \cref{fig:tiling}, 
\begin{equation}
  H_{\rm OBC} = -J\sum_{\braket{ij}} s_i s_j , \qquad s_i=\pm1 ,
  \label{eq:H_OBC}
\end{equation}
where $\braket{i, j}$ denotes a pair of nearest-neighbour sites on the $\{p,q\}$ tessellation,
$J>0$ is the uniform ferromagnetic coupling, and we set $J=1$ throughout. Keeping in mind the 
importance of boundary degrees of freedom, we study two limiting cases of boundary-boundary coupling. 

For a hyperbolic lattice with $\{p,q\}$ tessellation, all the bulk sites have the same coordination 
number $q$, while the sites in the outermost layer (boundary layer) $\partial V_N$ are connected only to 
spins in the same and inner layers and therefore have less than $q$  nearest neighbors. We simulate 
\cref{eq:H_OBC} on the full finite lattice (\cref{fig:tiling}), treating the bulk and boundary sites 
on the same footing. 

\begin{figure}[!tbh]
    \centering
    \includegraphics[width=\linewidth]{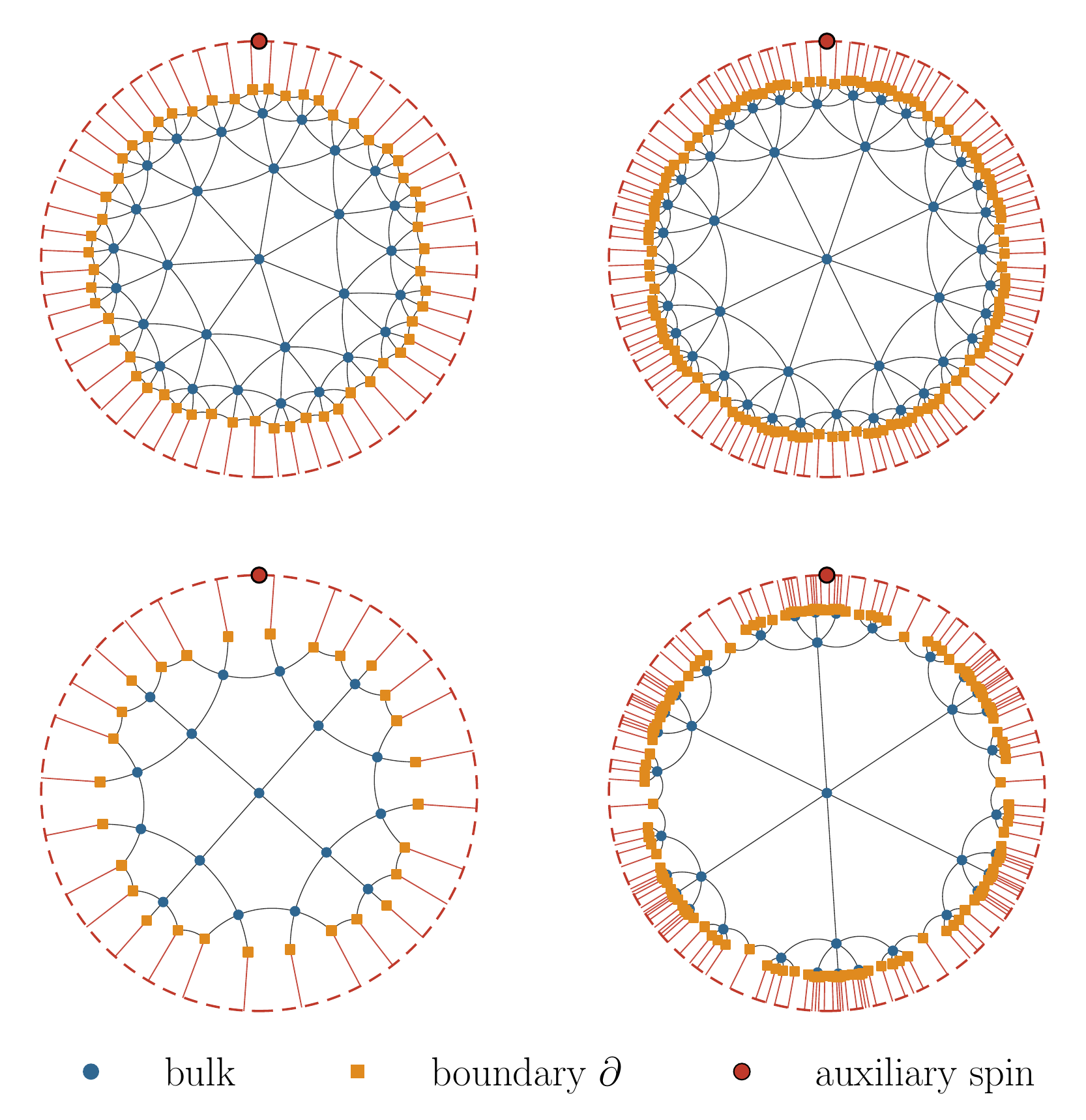}
    \caption{Wired boundary construction for the same four tilings. Every boundary spin is coupled 
    to a common auxiliary spin, shown by the dashed outer ring.}
    \label{fig:tiling-wbc}
\end{figure}

For the open boundary condition (OBC), the boundary spins are free to fluctuate, inducing a phase 
transition on the full lattice. We perform a finite-size scaling analysis in this framework in 
\cref{sec:OBC}. On the other hand, for the wired boundary condition (WBC), all boundary spins are 
coupled to a single \emph{auxiliary spin} $s_{\rm aux}=\pm1$ located in a fiducial layer outside 
the boundary layer of the lattice. This coupling naturally suppresses the independent fluctuations 
of the boundary spins. The Hamiltonian for this system is
\begin{equation}
  H_{\rm WBC} = H_{\rm OBC} - J_{b}s_{\rm{aux}} \sum_{i\in {\partial V_N}} s_i ,
  \label{eq:H-wbc}
\end{equation}
as illustrated in \cref{fig:tiling-wbc}. We consider $J_b = J =1$ throughout. The corresponding 
partition function is $Z_{\rm WBC}=\sum_{\{s\}} e^{-\beta H_{\rm WBC}}$. Since $s_{\rm aux}$ couples 
only to the boundary spins, the sum over $s_{\rm aux}$ can be carried out exactly in terms of the 
boundary magnetization $M_\partial=\sum_{i\in\partial V_N} s_i$,
\begin{equation}
\begin{split}
  Z_{\rm WBC} &= \sum_{\{s_i\}} e^{-\beta H_{\rm OBC}} \sum_{s_{\rm aux}=\pm1} e^{\beta J_b s_{\rm aux} M_\partial} \\
              &= \sum_{\{s_i\}} e^{-\beta H_{\rm OBC}}\, 2\cosh\!\left(\beta J_b M_\partial\right) ,
\end{split}
\label{eq:Z-wbc}
\end{equation}
where $\{s_i\}$ corresponds to the physical spins. The resulting factor depends on the total 
boundary magnetization, and thus induces an effective all-to-all coupling among the boundary spins.  
As discussed in \cite{f4sj-rwvj}, this suppresses boundary fluctuations and exposes bulk behaviour, 
which we discuss in \cref{sec:WBC}. 

An alternative way to control the boundary fluctuations is to apply an infinitesimal external 
magnetic field on the boundary sites,
\begin{equation}
    H_h = H_{\rm OBC} - h_\partial \sum_{i\in \partial V_N} s_i ,
  \label{eq:H_mag}
\end{equation}
where, $h_\partial$ denotes the boundary magnetic field. Unlike WBC, where the boundary spins couple 
through a common auxiliary spin which can flip (thus preserving the global $\mathbb{Z}_2$ symmetry), 
the infinitesimal field $h_\partial$ explicitly breaks the $\mathbb{Z}_2$ symmetry at the boundary 
and directly biases the boundary spins toward a preferred orientation. Subsequently, it provides a 
direct probe of boundary-induced ordering and allows one to distinguish it from spontaneous ordering 
of the bulk. This distinction is central to the boundary-induced intermediate \emph{Eggarter} phase 
\cite{f4sj-rwvj}, discussed in \cref{sec:eggarter}. The magnetization is defined using the physical 
spins, 
\begin{equation}\label{eq:mag}
    m = \frac{1}{N} \sum_{i=1}^N s_i ,
\end{equation}
where $N$ counts the physical spins (excluding the auxiliary spin). The susceptibility $\chi$ 
is defined as 
\begin{equation}\label{eq:chi_metro}
    \chi = N \braket{ m^2 }.
\end{equation}
In numerical simulations, especially when local algorithms are being used, it is more conventional to 
define the connected $\chi_c$ by subtracting the absolute value of the magnetization,
\begin{equation}
    \chi_c = N\left(\braket{m^2} - \braket{|m|}^2\right). 
\end{equation}
We will also have occasion to present results of the Binder cumulant $U_2$, 
\begin{equation}
    U_2 = 1 - \frac{\braket{m^4}}{3\braket{m^2}^2},
    \label{eq:binder}
\end{equation}
which provides a renormalized four-point coupling. The mean energy is obtained as,
\begin{equation}
  \braket{E} =  -\frac{\partial\ln Z}{\partial\beta} = -J\sum_{\braket{ij}}  \braket{s_i s_j},
  \label{eq:metro_E}
\end{equation} 
where ${\braket{ij}}$ denotes nearest neighbor pairs. We have studied the Ising model for both 
boundary conditions, as well as explored the effect of an infinitesimal external magnetic field 
on the boundary using Markov chain Monte Carlo (MCMC) methods, as described in \cref{sec:algorithm}.

\subsection{Algorithms and Benchmarks}\label{sec:algorithm}
 The numerical methods used to simulate the Ising model on a finite $\{p,q\}$ tessellation employ two 
different categories of algorithms: the single-spin Metropolis update and the global-update worm algorithm 
\cite{Prokof_ev_2001, Wolff2009, PhysRevLett.99.110601}.
The Metropolis algorithm updates the spin configurations directly and serves as an independent 
benchmark for the worm algorithm. It operates as follows: an attempt is made to flip each spin of the 
configuration in turn. If the total energy is lowered by the spin flip, it is immediately accepted, 
while it is accepted with a probability ${\rm exp}(-\beta \Delta E)$ otherwise, where 
$\Delta E = E_{\rm final} - E_{\rm initial}$. This satisfies a detailed balance condition, and guarantees 
that the algorithm generates configurations from the equilibrium distribution ${\rm exp}(-\beta H)$. 
Since one spin is updated at a time, it typically gives rise to large auto-correlations when the 
spins display collective behaviour. This is 
well-known as critical slowing near a second-order phase transition. The worm algorithm, in contrast,
works on a basis of bonds connecting nearest-neighbour sites obtained by integrating out the spin
variables using a \emph{high-temperature series expansion} of the partition function. The algorithm,
described below, is able to perform collective loop updates and thus can address 
critical slowing down.

 We now discuss the worm algorithm, which samples bond configurations arising from the 
high-temperature expansion of the Ising model. Consider OBC with a uniform coupling $J$, and use
 \begin{equation}
    e^{\beta J s_i s_j} = \cosh(\beta J) \sum_{n_{ij}=0,1}(t\,s_i s_j)^{n_{ij}},~~t=\tanh(\beta J),
 \end{equation}
we introduce an occupation variable $n_{ij}\in\{0,1\}$ on each bond. A bond is occupied when 
$n_{ij}=1$. For a given bond configuration, the spin at site $i$ appears with power $k_i=\sum_j n_{ij}$, and
 \begin{equation}
  \sum_{s_i=\pm1}s_i^{k_i} = \begin{cases}
                               2, & k_i \text{ even},\\
                               0, & k_i \text{ odd}.
                             \end{cases}
 \end{equation}
 Therefore, the partition function becomes
 \begin{equation}
 \begin{split}
   Z &= 2^{N}\left(\cosh\beta J\right)^{M}\sum_{\{n\}} t^{|A|} \prod_{i} \delta_{k_i,\,{\rm even}} \\
     &= 2^{N}\left(\cosh\beta J\right)^{M}\sum_{\partial A=\emptyset} t^{|A|} ,
 \end{split}
 \label{eq:HT_Z}
 \end{equation}
where $A$ is the set of occupied bonds, $|A| = \sum_{\braket{ij}} n_{ij}$ is the total number 
of occupied bonds, and $\partial A$ is the set of sites with odd $k_i$, and $M$ the number of 
bonds of the lattice. Thus, $Z$ receives contributions from only even-degree subgraphs (i.e., 
those with $\partial A=\emptyset$). These can be decomposed into non-disjoint closed loops. 

\begin{figure}[!tbh]
    \centering
    \includegraphics[width=\linewidth]{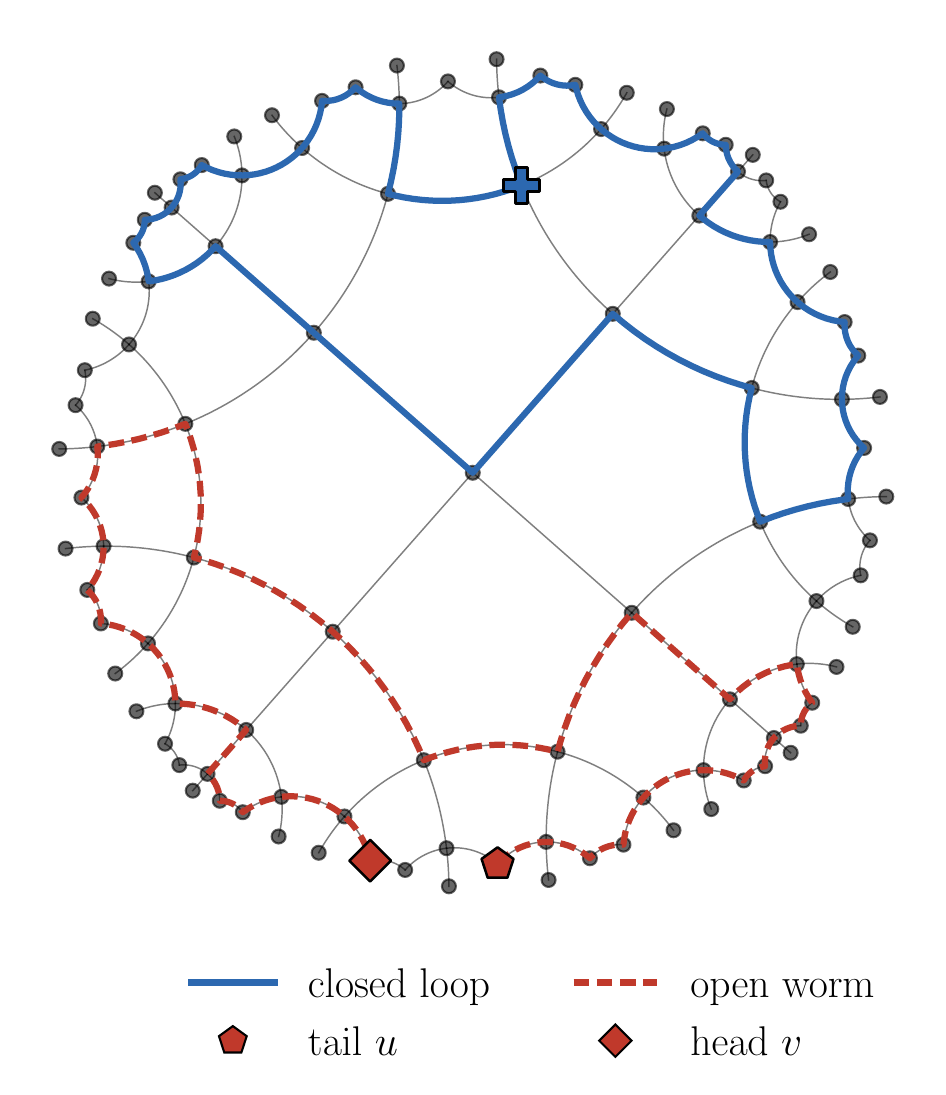}
    \caption{Two configurations generated by worm algorithm on the $\{5, 4\}$ tessellation with OBC. 
    The solid line represents a closed loop, formed when the head $v$ meets the tail $u$ at 
    $\meetsym$ marker, while the dashed line describes a open configuration, with the tail 
    $u$ and the head $v$ indicated by a pentagon ($\tailsym$) and diamond ($\headsym$) symbol, 
    respectively.}
    \label{fig:worm_path}
 \end{figure}

  The worm algorithm considers a larger ensemble by enlarging this configuration space with 
two marked vertices, the tail $u$ and the head $v$ (these can be thought of as sources and
sinks, or defects). For $u\ne v$, these are the only vertices with odd occupied degree; when 
$u=v$, all vertices have even occupied degree. We denote the corresponding defect set by
 \begin{equation}
   D(u,v)= \begin{cases}
              \{u,v\},   & u\ne v,\\
              \emptyset, & u=v.
           \end{cases}
 \end{equation}
 The two-point function is then
 \begin{equation}
  \braket{s_u s_v} = \frac{\displaystyle \sum_{\partial A=D(u,v)}t^{|A|}}{\displaystyle \sum_{\partial A=\emptyset}t^{|A|}}.
  \label{eq:worm_G}
 \end{equation}
 An open configuration can be decomposed into a path connecting $u$ and $v$, together with closed loops; 
this decomposition is generally not unique.  A schematic representation of an open and closed 
configuration sampled by the worm update on the $\{5,4\}$ tessellation with OBC is depicted in \cref{fig:worm_path}.
 The extended configurations $X=(A,u,v)$ are sampled with weight
 \begin{equation}\label{eq:worm_weight}
   \pi(X)\propto t^{|A|}\,\mathbf{1}_{\partial A=D(u,v)}. 
 \end{equation}
 A worm starts with coincident endpoints $u=v$ at a randomly chosen vertex. Keeping the tail fixed, 
we propose to move the head from $i=v$ to a neighbouring site $j$ and simultaneously toggle the 
intervening bond, $n_{ij}\to1-n_{ij}$. The combined move preserves the defect constraint. For a 
proposed transition $X\to X'$, a detailed balance requires
 \begin{align}
   & \pi(X)P_{\rm prop}(X\to X')P_{\rm acc}(X\to X') = \nonumber \\
    &~~~~~~~~\pi(X')P_{\rm prop}(X'\to X)P_{\rm acc}(X'\to X).
 \end{align}
 Here, $P_{\rm prop}(X\to X')$ indicates the probability to make a proposal to change the configuration
from $X \to X'$ and $P_{\rm acc}(X\to X')$ is the probability to accept the proposal. 
This is satisfied by the following Metropolis-Hastings acceptance probability
 \begin{equation}
  P_{\rm acc}(X\to X') = \min\!\left[ 1,\, \frac{\pi(X')P_{\rm prop}(X'\to X)}{\pi(X)P_{\rm prop}(X\to X')} \right].
 \end{equation} 

  For OBC, the head chooses uniformly among $q+1$ options: $q$ neighbour slots and one 
stay-put option. At a boundary site, slots corresponding to missing neighbours generate null moves.
The proposal probability along every existing bond is therefore symmetric,
$P_{\rm prop}(i\to j)=P_{\rm prop}(j\to i)=1/(q+1)$. 
Since toggling a bond changes $|A|$ by $1-2n_{ij}$, the acceptance probability becomes
  \begin{equation}
    P_{\rm acc}(X\to X') = \min\!\left(1,t^{\,1-2n_{ij}}\right),
    \label{eq:worm_acc_obc}
  \end{equation}
where $n_{ij}$ is the occupation before the move. Accepted moves update both the bond and the head position; 
rejected or null moves leave the configuration unchanged. When the head returns to the tail, the worm closes 
and the bond configuration again contributes to the partition function. A new worm can then be started by 
relocating both endpoints to a randomly chosen vertex without changing the bonds.
For the OBC at the boundaries, the worm is offered the same number of choices as a bulk site, but it is
not displaced if the new site is not another boundary or a bulk site.
 
  For WBC, the graph includes the auxiliary vertex and its bonds to the $N_b=|\partial V_N|$ boundary sites. 
The bond weight is now
 \begin{equation}
    w(A)=\prod_{\langle ij\rangle\in A}t_{ij},~~~t_{ij}=\tanh(\beta J_{ij}),
 \end{equation}
with $J_{ij}=J_b$ on auxiliary bonds and $J_{ij}=J$ otherwise. The corresponding partition function is
\begin{equation}
   Z_{\rm WBC} = 2^{N+1} \prod_{\langle ij\rangle\in\mathcal E_{\rm WBC}} \cosh(\beta J_{ij})
     \sum_{\partial A=\emptyset}w(A),
\end{equation}
where $\mathcal E_{\rm WBC}$ includes both physical and auxiliary bonds, and $N$ continues to count 
physical spins only.

 For the WBC update, the head at vertex $i$ chooses uniformly among its $d_i$ neighbours and one 
stay-put option, giving $D_i=d_i+1$ choices. Here $d_i$ includes the auxiliary bond for boundary vertices, 
and $d_{\rm aux}=N_b$. Consequently, $P_{\rm prop}(i\to j)=1/D_i$, and detailed balance gives
\begin{equation}
     P_{\rm acc}(X\to X') = \min\!\left( 1,\, t_{ij}^{\,1-2n_{ij}}\frac{D_i}{D_j} \right).
     \label{eq:worm_acc_wbc}
\end{equation}
 
  For OBC, the reduced susceptibility is related to the probability of coincident worm endpoints by 
\cite{Wolff2009}
\begin{equation}
    \chi = N\braket{m^2} = \frac{1}{\langle\!\langle\delta_{u,v}\rangle\!\rangle},
  \label{eq:worm_chi_obc}
\end{equation}
where double brackets denote averages in the extended ensemble consisting of the heads and tails
as defined earlier. These averages include both open and closed configurations, with their equilibrium 
sampling weights, rather than only completed worms.

 During the actual simulations involving WBCs, we used a re-weighting factor in order to bias the worm 
to update the auxiliary and the boundary spins more frequently. Otherwise, the worm head often gets 
stuck at the auxiliary spin. Furthermore, one has to measure the physical susceptibility when 
simulating the full WBC graph by discarding auxiliary-spin contributions. This can be done by defining
a new observable $I$, which is $I_{\rm phys}(i)=1$ for a physical spin vertex and zero for the auxiliary 
spin vertex. Thus, the appropriate estimator for susceptibility automatically selects the physical sector,
 \begin{equation}
   \chi = \frac{\langle\!\langle I_{\rm phys}(u)I_{\rm phys}(v) \rangle\!\rangle}
               {\langle\!\langle \delta_{u,v}I_{\rm phys}(u) \rangle\!\rangle}.
          \label{eq:worm_chi_wbc}
 \end{equation}
 For the same unbiased endpoint ensemble, the central-auxiliary correlator is
 \begin{equation}
   \braket{s_0 s_{\rm aux}} = \frac{\langle\!\langle \delta_{u,0}\delta_{v,{\rm aux}} \rangle\!\rangle}
       {\langle\!\langle \delta_{u,0}\delta_{v,0} \rangle\!\rangle},
       \label{eq:worm_central}
 \end{equation}
which we use in order to show critical behaviour on lattices with WBC. The energy is obtained by 
differentiating the partition function with respect to $\beta$. For OBC with uniform coupling,
 \begin{equation}
   \braket{E} = -\frac{\partial\ln Z}{\partial\beta} 
              = -J\left[ Mt + \left(t^{-1}-t\right) \braket{|A|}_{\rm closed}\right],
   \label{eq:worm_E_obc}
 \end{equation}
where the average is conditioned on $u=v$. More generally, including unequal auxiliary-bond couplings,
 \begin{equation}\label{eq:worm_E_wbc}
   \braket{E} =-\sum_{\braket{ij}} J_{ij} \left[ t_{ij} + \left(t_{ij}^{-1}-t_{ij}\right)
                 \braket{n_{ij}}_{\rm closed}\right].
\end{equation}
 Ergodicity, together with the detailed balance guarantees that both the Metropolis and worm 
updates converge to their respective unique equilibrium distribution, independently of the 
initial configuration. Starting from an initial probability distribution $\pi_0(C)$ over the 
configuration $C$, each update step generates a new distribution $\pi_k(C)$, and the sequence 
converges to the stationary equilibrium distribution $\pi(C)$.  For the Metropolis update this
is the Boltzmann weight $\pi(C)\propto e^{-\beta H(C)}$, while for the worm algorithm it is the
weight $\pi(X)$ of the extended configurations in \cref{eq:worm_weight}.

 Once the system has equilibrated, physical operators can be measured on configurations. 
 We denote the configuration at the $m$-th Monte Carlo time by $C_m$, and the measurement of an 
observable $\mathcal{O}$ on it by $\mathcal{O}_m\equiv\mathcal{O}(C_m)$. Depending on the algorithm 
used, consecutive configurations $C_m$ and $C_{m+1}$ are generally correlated, since each new 
configuration is generated using from the previous one, instead of being sampled independently 
from the equilibrium distribution. Therefore, the measurements of an observable $\mathcal{O}_m$ 
and $\mathcal{O}_{m+1}$ are also autocorrelated. Near a critical point, where systems exhibit 
long-range order, more Monte Carlo updates are required to remove autocorrelations, particularly 
for local update algorithms. This is known as \emph{critical slowing down}, and affects both the 
equilibration time, as well as the number of independent measurements in Monte Carlo chain. The 
computation of the \emph{autocorrelation function} and the associated \emph{autocorrelation time} 
is required to make the statistical errors precise \cite{montecarlo_stat, texbook, morningstar2007monte, 
WOLFF2004143, Banerjee_2021, Joswig_2023}. 

In equilibrium, the sequence of measurements $\{\mathcal{O}_m\}$ constitutes a stationary stochastic 
process, and $\braket{\mathcal{O}}$ is estimated by the sample mean 
$\overline{\mathcal{O}}=N_{\rm conf}^{-1}\sum_m\mathcal{O}_m$ over $N_{\rm conf}$ measurements. By the
central limit theorem, $\overline{\mathcal{O}}$ is asymptotically Gaussian about
$\braket{\mathcal{O}}$,  with $\braket{\mathcal{O}}=\overline{\mathcal{O}}\pm\sigma_{\overline{\mathcal{O}}}$.
For statistically independent measurements, the unbiased estimator of its variance is
\begin{equation}
  \sigma^2_{\overline{\mathcal{O}}} = \frac{1}{N_{\rm conf}(N_{\rm conf}-1)}
  \sum_{m=1}^{N_{\rm conf}}\left(\mathcal{O}_m-\overline{\mathcal{O}}\right)^2 .
  \label{eq:naive_error}
\end{equation}
As argued before, measurements at different MC times are generally correlated, leading to a non-vanishing
autocorrelation function. For large $N_{\rm conf}$
\begin{equation}
\begin{split}
    \sigma^2_{\overline{\mathcal{O}}} 
  &= \frac{1}{N_{\rm conf}^2} \sum_{m, n=1}^{N_{\rm conf}} 
      \left(\mathcal{O}_m-\overline{\mathcal{O}}\right)\left(\mathcal{O}_n-\overline{\mathcal{O}}\right)\\
  &= \frac{1}{N_{\rm conf}^2}\sum_{m,n=1}^{N_{\rm conf}}\Gamma_{\mathcal{O}}(m-n) ,
  \end{split}
  \label{eq:corr_error}
\end{equation}
where, the autocorrelation function between measurements at $m$-th and $(m+\tau)$-th times, is described by
\begin{equation}
    \Gamma_{\mathcal{O}}(\tau) = \braket{\mathcal{O}_m\,\mathcal{O}_{m+\tau}} - \braket{\mathcal{O}}^2
  = \Gamma_{\mathcal{O}}(-\tau) \;\stackrel{\tau\gg1}{\sim}\; C_0\, e^{-\tau/\tau_{\rm exp}}.
  \label{eq:autocorr}
\end{equation}
The autocorrelation function $\Gamma_{\mathcal{O}}(\tau)$ depends only on the separation $\tau$, 
and characterizes the Monte Carlo dynamics rather than the equilibrium distribution. Its asymptotic 
decay defines the \emph{autocorrelation time} $\tau_{\rm exp}$, which provides an estimate about 
how long it takes configurations to decorrelate. A more relevant measure is the 
\emph{integrated autocorrelation time}, which accounts for the cumulative contribution of correlations 
at different time separations. Defining the normalized autocorrelation function as
\begin{equation}
 \rho_{\mathcal{O}}(\tau) = \frac{\Gamma_{\mathcal{O}}(\tau)}{\Gamma_{\mathcal{O}}(0)},
\end{equation}
The integrated autocorrelation time is given by,
\begin{equation}
  \tau_{{\rm int},\mathcal O} = \frac12+\sum_{\tau=1}^{\infty}\rho_{\mathcal O}(\tau),
\end{equation}
and for finite Monte Carlo samples is computed using
\begin{equation}
  2\tau_{\rm int} \approx \sum_{r=-W}^{W}\rho_{\mathcal{O}}(\tau)
  = 1 + 2\sum_{\tau=1}^{W}\rho_{\mathcal{O}}(\tau),
  \label{eq:tauint}
\end{equation}
where, $W$ is the cutoff in Monte Carlo time required for a finite data set. Up to corrections of 
order $\tau_{\rm exp}/N_{\rm conf}$, \cref{eq:corr_error} then becomes
\begin{equation}
  \sigma^2_{\overline{\mathcal{O}}} = \frac{2\tau_{\rm int}\,\Gamma_{\mathcal{O}}(0)}{N_{\rm conf}}
  = \frac{\Gamma_{\mathcal{O}}(0)}{N_{\rm eff}}, \qquad N_{\rm eff}=\frac{N_{\rm conf}}{2\tau_{\rm int}} .
  \label{eq:error_tauint}
\end{equation}
Thus, the statistical precision obtained from $N_{\rm conf}$ correlated measurements is equivalent 
to that obtained from $N_{\rm eff}$ statistically independent measurements. Typically, the
integrated autocorrelation time depends both on the observable and the update algorithm. 
Near a critical point, the autocorrelation times grow with the correlation length as $\tau\sim\xi^{z}$ 
\cite{montecarlo_stat, texbook}, where $z$ is the dynamical critical exponent. Accordingly, the exponent 
$z_{{\rm int},\mathcal{O}}$ associated with $\tau_{\rm int}$ differs between observables, and is bounded 
from above by $z_{\rm exp}$ corresponding to the slowest-updated observable. On a lattice in a flat geometry
with linear size $L$, the correlation length $\xi$ is bounded by $L$ at criticality, giving 
$\tau_{\rm int}\sim L^{z}$. In the literature, Metropolis updates are known to exhibit critical slowing down, 
characterized by a dynamical critical exponent typically of order 
$z\approx 2$ \cite{Janke2008, RevModPhys.49.435}. 

For hyperbolic lattices, we characterize the size dependence of the integrated autocorrelation time
$\tau_{\rm int}$ in terms of the volume $N$ as, $\tau_{\rm int} = A N^{z_N}$. where $A_{\mathcal{O}}$ 
is an amplitude and $z_{N,\mathcal{O}}$ is the corresponding volume-scaling exponent. We use this 
exponent to compare the dynamical efficiency of the Metropolis and worm updates.
\begin{figure}[!tbh]
    \centering
    \includegraphics[width=\linewidth]{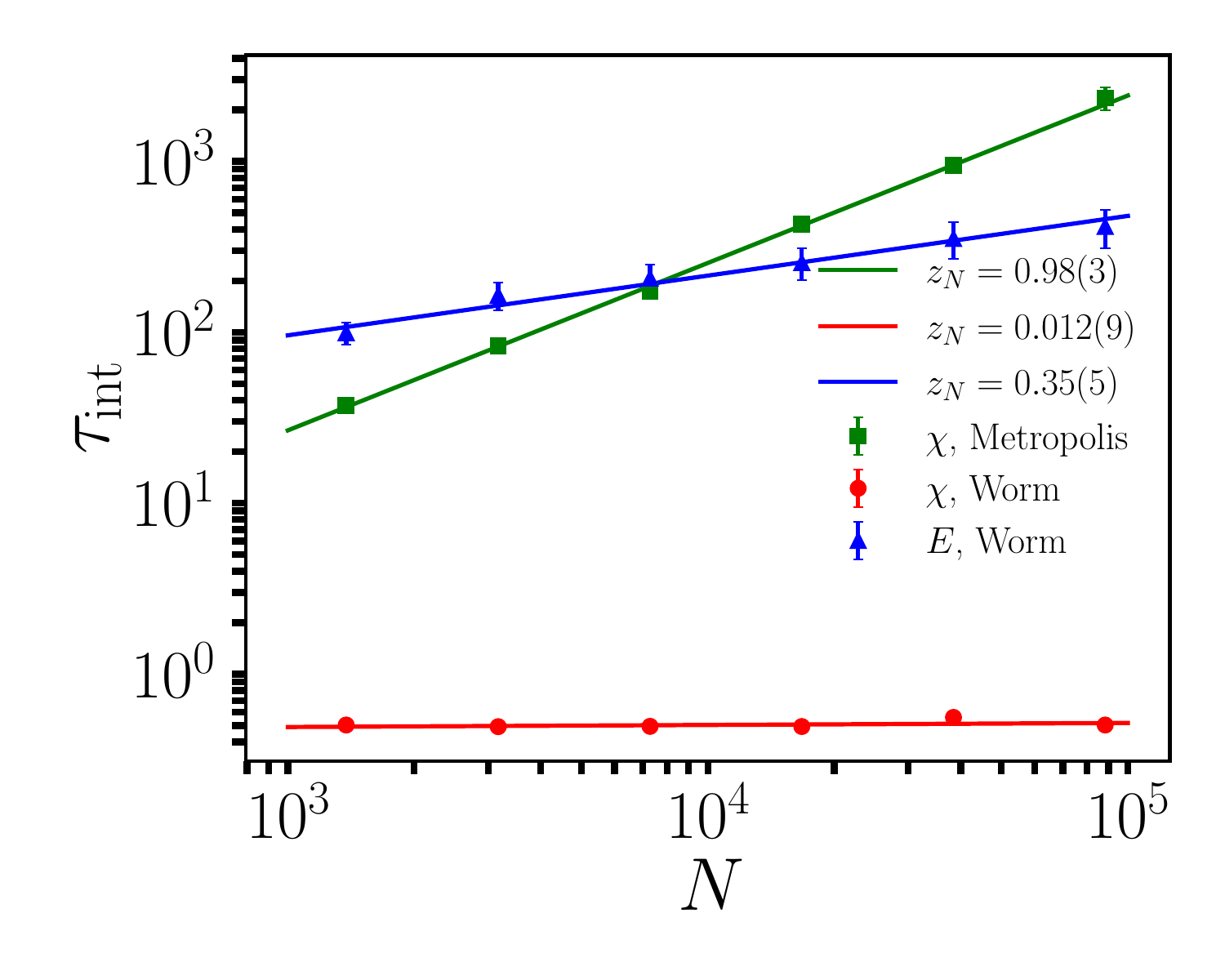}
    \caption{Integrated autocorrelation time as a function of total number of sites $N$ 
    at critical point $\beta_c = 0.631$ for the $\{5,4\}$ lattice. Monte Carlo time is measured 
    in units of one sweep (comprising $N$ attempted spin updates) for the Metropolis updates and 
    of one completed worm for the worm algorithm.}
    \label{fig:tau_int}
\end{figure}
 A comparison of dynamic exponents $z_N$ for both updates is presented in the \cref{fig:tau_int} 
at the transition $\beta_c = 0.631$ for the $\{5,4\}$ tessellation. Integrated autocorrelation 
times and corresponding statistical errors are estimated using the $\Gamma$-method \cite{WOLFF2004143}, 
implemented in the package \texttt{py-uwerr} \cite{Hesse2024pyuwerr}. For $\chi$, Metropolis update 
exhibits a growth of $\tau_{\rm int}$ with $N$, with $z_N = 0.98(3)$, $\chi^2/\rm{d.o.f} = 0.31$. 
In contrast, the worm algorithm yields $z_N=0.012(9)$ with $\chi^2/\rm{d.o.f} = 2.2$ for the same 
observable, indicating a nearly size-independent autocorrelation time. For the energy $E$, we obtain 
$z_N=0.35(5),  \chi^2/\rm{d.o.f} = 0.28$ from the worm algorithm. The weaker growth of 
$\tau_{\rm int}$ for the worm updates demonstrates that it can update the system more efficiently 
at the critical regime.

\begin{table}[!tbh] 
    \centering
    \renewcommand{\arraystretch}{1.7}
    \resizebox{0.5\textwidth}{!}{
    \begin{tabular}{c c cc cc}
        \hline
        $N$ & $\beta$ & $\chi_{\rm Metro}$ & $\chi_{\rm worm}$& $\braket{E}_{\rm Metro}/N$ & $\braket{E}_{\rm worm}/N$ \\
        \hline
        1377  & 0.635 & 173.99(12)  & 173.984(45) & $-0.88306(1)$ & $-0.88305(1)$  \\
              & 0.80  & 729.13(42)  & 728.94(22)  & $-1.0846(2)$ & $-1.08460(1)$  \\
        \hline
        3169  & 0.63  & 293.39(14)  & 293.20(15)  & $-0.877647(3)$  & $-0.87766(2)$ \\
        \hline
        7285 & 0.68  & 1326.32(93) & 1325.00(13)  & $-0.950834(3)$  & $-0.95084(2)$ \\
             & 0.70  & 1751.7(24)  & 1750(2)  & $-0.977771(6)$  & $-0.97776(2)$ \\
        \hline
    \end{tabular}
    }
    \caption{Benchmark of the observables $\chi_{\rm worm}$(\cref{eq:worm_chi_obc}) and  
    $\braket{E_{\rm worm}}/N$ (\cref{eq:worm_E_obc}) from the worm algorithm against 
    $\chi_{\rm Metro}$(\cref{eq:chi_metro}) and  $\braket{E_{\rm Metro}}/N$ (\cref{eq:metro_E}) 
    from the Metropolis update, on the $\{5,4\}$ tessellation with OBC.} 
    \label{tab:obc_benchmark}
\end{table}

To benchmark the worm algorithm, we compare its estimates of the susceptibility $\chi$ and energy 
$E$ with those obtained from the Metropolis updates. The results for OBC are presented in 
\cref{tab:obc_benchmark} for several system sizes and couplings $\beta$ for $\{5,4\}$ tessellation. 
For both observables, the two algorithms agree within statistical errors.
\begin{table}[!tbh]
    \centering
    \renewcommand{\arraystretch}{1.7}
    \resizebox{0.5\textwidth}{!}{
    \begin{tabular}{c c cc cc}
        \hline
        $N$ & $\beta$ & $\chi_{\rm Metro}$ & $\chi_{\rm worm}$ & $\braket{E}_{\rm Metro}/N$ & $\braket{E}_{\rm worm}/N$ \\
        \hline
         257   & 0.35  & 64.356(3)   & 64.41(11)    & $-0.89735(2)$  & $-0.89735(33)$ \\
              & 0.36  & 71.630(3)   & 71.68(11)    & $-0.94077(2)$  & $-0.94089(21)$ \\
        \hline
        597    & 0.35  & 143.934(5)  & 144.04(30)   &  $-0.89640(1)$&  $-0.89649(24)$\\
              & 0.36  & 161.126(5)  & 161.24(31)   & $-0.94023(1)$  & $-0.94019(27)$ \\
        \hline
         1377     & 0.35  & 326.149(9)  & 327(1)     &  $-0.89571(1)$  &$-0.89557(19)  $\\
              & 0.36  & 366.105(9)  & 366.05(82)   & $-0.93973(1)$  & $-0.93970(13)$ \\
        \hline
    \end{tabular}
    }
    \caption{Comparison of  $\chi_{\rm worm}$(\cref{eq:worm_chi_wbc}) and  $\braket{E_{\rm worm}}/N$ 
    (\cref{eq:worm_E_wbc}) obtained with the worm algorithm, with $\chi_{\rm Metro}$(\cref{eq:chi_metro}) 
    and  $\braket{E_{\rm Metro}}/N$ (\cref{eq:metro_E}), obtained from the Metropolis update, on the 
    $\{5,4\}$ tessellation with WBC.} 
    \label{tab:wbc_benchmark}
\end{table}

A similar comparison for WBC is shown in \cref{tab:wbc_benchmark} for $\{5,4\}$ tessellations. Again, 
the susceptibility $\chi$ and the energy $E$ obtained from the two algorithms agree within statistical 
errors for all system sizes and inverse temperatures considered. These benchmarks provide an independent 
validation of the worm algorithm and of the estimators derived for these hyperbolic lattices.

\section{Results}\label{sec:results}
\subsection{Universality with OBC}\label{sec:OBC}
We first study the Ising model with OBC on finite $\{p,q\}$ tessellations. Our aim is to examine 
whether the transition of the full finite lattice is characterized by the same universal behavior 
on general $\{p,q\}$ tessellations, which we investigate with finite-size scaling (FSS) for the 
$\{3,7\}$, $\{3,8\}$, $\{5,4\}$ and $\{6,6\}$ tessellations.

\begin{figure*}[!tbh]
    \centering
    \includegraphics[width=0.48\linewidth]{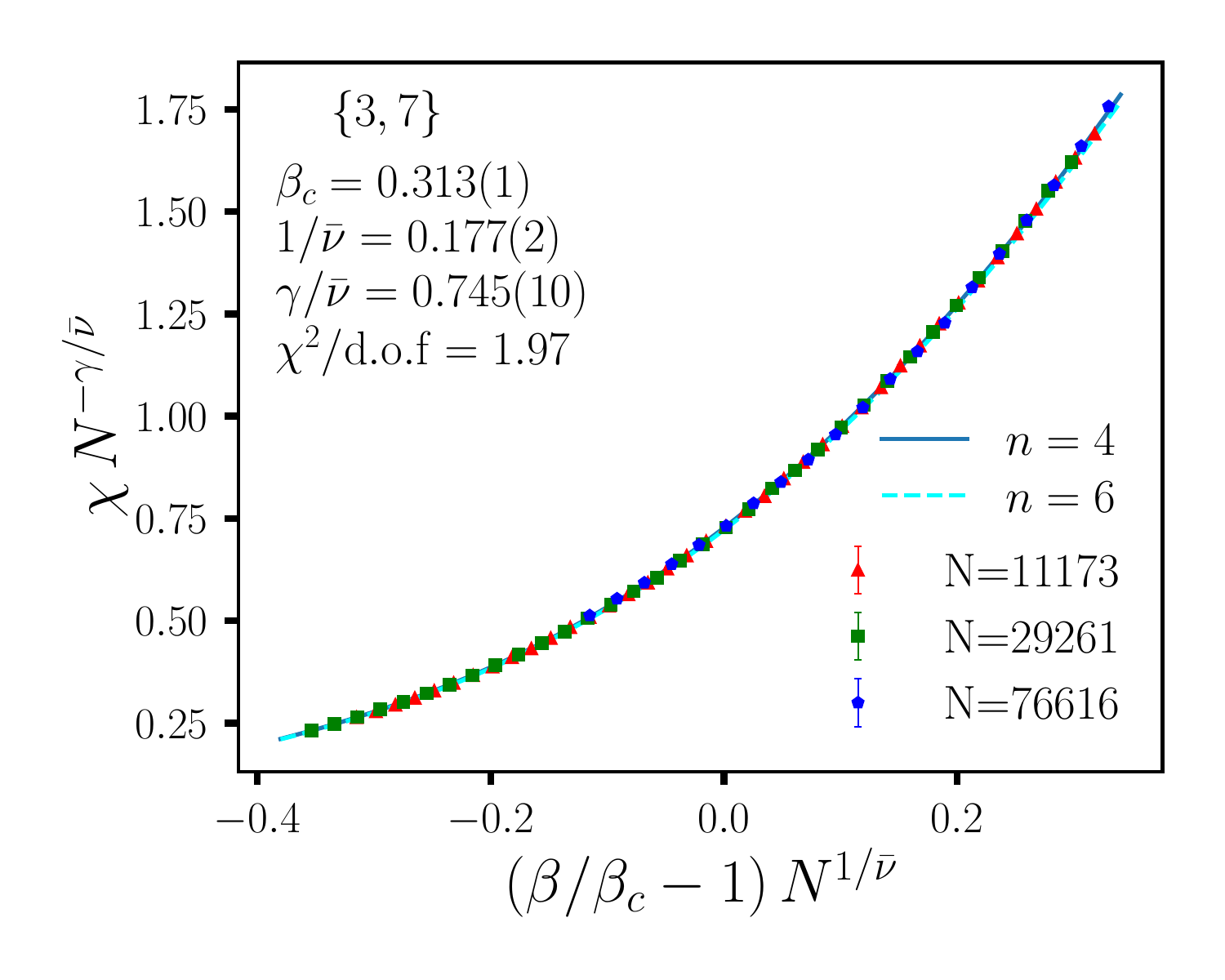}
    \includegraphics[width=0.48\linewidth]{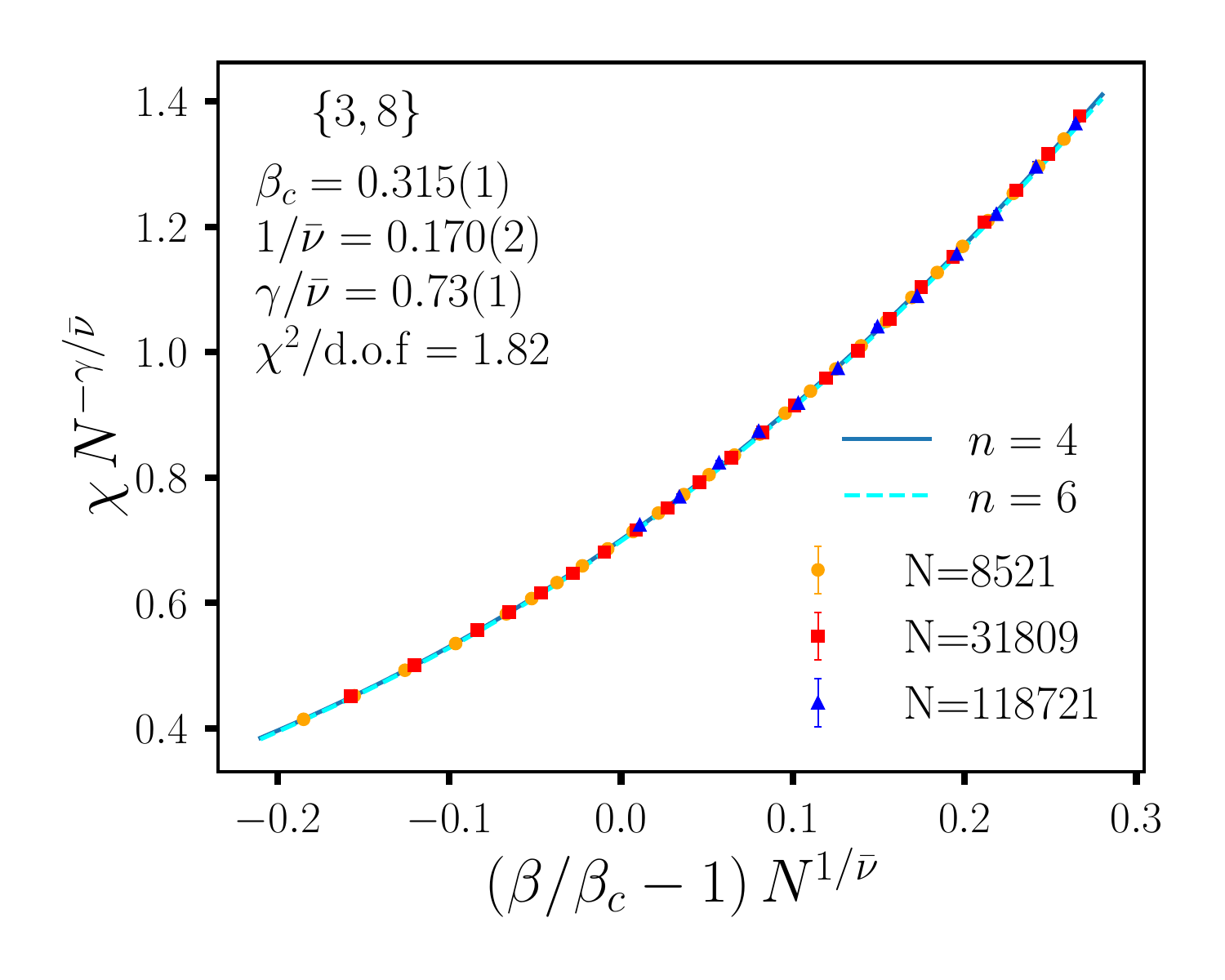}
    \includegraphics[width=0.48\linewidth]{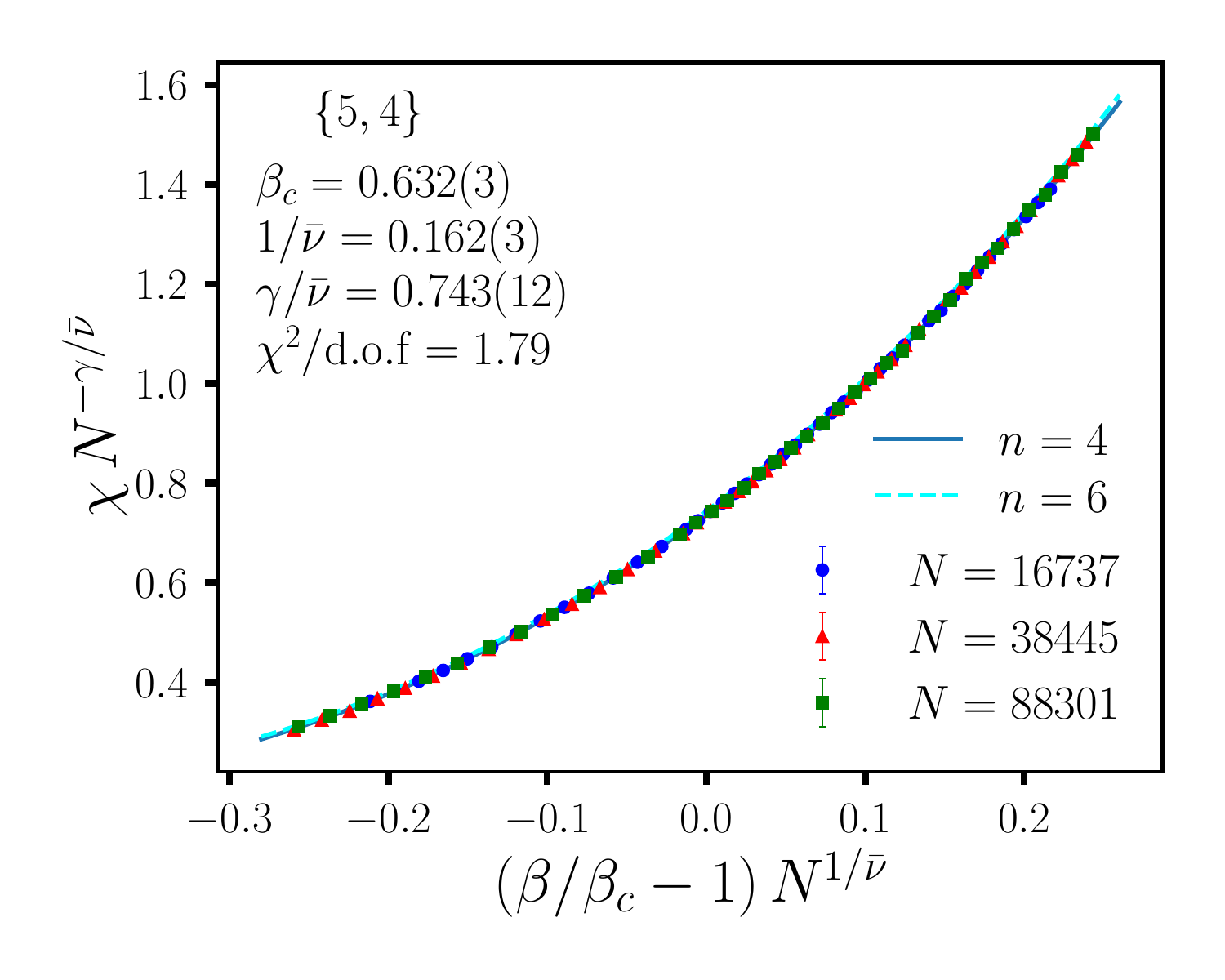}
    \includegraphics[width=0.48\linewidth]{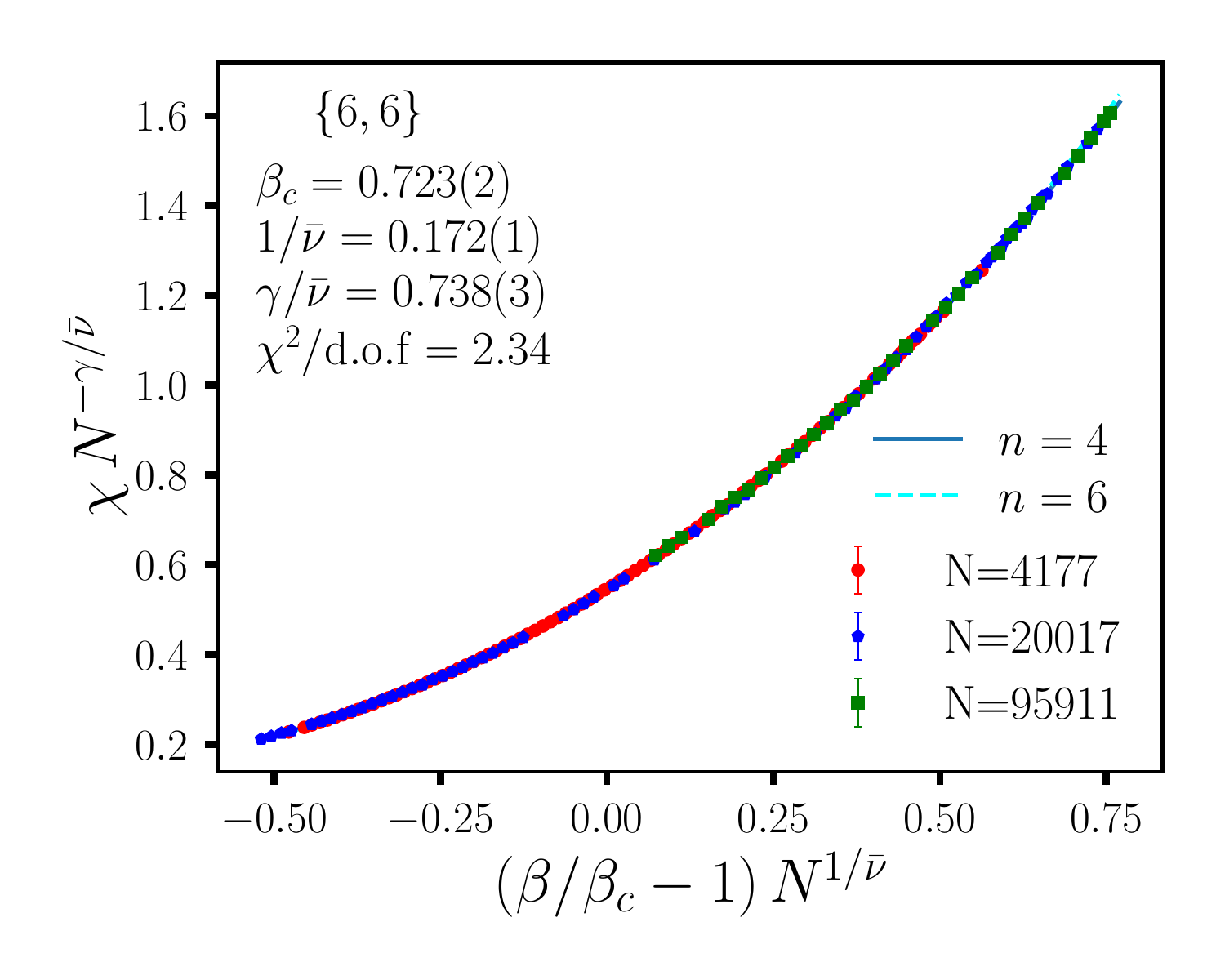}
    \caption{FSS collapse of the susceptibility with OBC for the $\{3,7\}$ (top left),
    $\{3,8\}$ (top right), $\{5,4\}$ (bottom left) and $\{6,6\}$ (bottom right)
    tessellations, using volume scaling with the total number of sites $N$. Solid
    and dashed lines are fits to fourth- and sixth-order scaling polynomials; the
    quoted exponents use the quartic fit.}
    \label{fig:FSS_OBC}
\end{figure*}

 As argued before, while on flat lattices, the system size $L$ is typically used as the scale
in FSS studies, for hyperbolic lattices it is not a-priori clear if it continues to be the
relevant scale. Specifically, our attempts to use $L$ in the FSS studies make it non-trivial to
identify critical exponents and yield an inconsistent critical point, as already observed in \cite{Shima_2006}.
We instead consider the total number
of sites $N$ as the scaling variable. Volume-based FSS has a precedent in \cite{Kulkarni2022}, 
who found that critical rounding for the Ising model on random regular graphs is governed by 
$tN^{1/2}$ rather than a power of the graph diameter (which grows logarithmically with $N$,
not with $L$). Hyperbolic tessellations likewise exhibit exponential volume growth with radial extent, 
motivating us to test a scaling variable $tN^{1/\bar{\nu}}$ for the susceptibility of the full open 
system. Moreover, the usual scaling hypothesis is still necessary to determine the critical exponents.
We therefore determined $\bar{\nu}$ numerically and assess the volume-based scaling hypothesis 
through data collapse and its stability against finite-size corrections.

 The usual assumption that at the vicinity of the critical point, the susceptibility of finite 
systems obeys the scaling behaviour still holds, and we use the ansatz,
\begin{equation}
    \chi(t,N) = N^{\gamma/\bar\nu} F(t N^{1/\bar\nu}), \quad t = (\beta/\beta_c -1). \label{eq:FSS_chi}
\end{equation}
 Close to the transition, the scaling variable $x=t\,N^{1/\bar\nu}$ is small, and the scaling 
function \cref{eq:FSS_chi} can be expanded as a polynomial. The scaling function $F(x)$ then 
takes the form,
\begin{equation}
  F(x)= a_0 + a_1x + a_2x^2 + a_3 x^3 + \cdots . \label{eq:FSS_fit}
\end{equation}
We truncated the polynomial fit to $n=4$ and $n=6$ in our work. For each tessellation, we fit 
\cref{eq:FSS_chi} simultaneously to the susceptibility data of the largest lattices by non-linear 
least squares fitting to obtain the critical exponents $\gamma/\bar \nu$, $1/\bar\nu$ and $\beta_c$ 
as optimal fitting parameters. The representative scaling collapse for all four tessellations along with 
the susceptibility data is depicted in \cref{fig:FSS_OBC}. The fit parameters obtained are summarized 
in \cref{tab:fss_exponents} for all four lattices. The critical point varies monotonically from 
$0.313(1)$ for the $\{3,7\}$ to $0.723(2)$ for $\{6,6\}$ tessellation, as a function of the curvature. 
This variation is natural given the critical value between the ferromagnetic and the paramagnetic phase is
model dependent, and the local connectivity changes between the tessellations. This is also observed
for models on flat space, where different microscopic models with the same global symmetries can have 
different critical points. 

 The hallmark of a universality class is rather the set of critical exponents, which depend on the
symmetries, the space-time dimensions and possibly the curvature, but is agnostic about the microscopic 
description of the model. In our case, indeed, we see that the exponent $\gamma/\bar{\nu} \sim 0.73-0.74$ 
and $1/\bar{\nu} = 0.170 - 0.177$ across the different tessellations, with one value at $0.162(3)$, 
and hinting at the existence of a new universality class. The Ising model on the flat $d=2$ instead
has $\nu=1$ and $\gamma/\nu=7/4$, while the corresponding value for $d=\infty$ (corresponding to mean
field) is $\nu=1/2$ and $\gamma/\nu=2$. Our values are significantly different from any of these values
and also different from the values in other dimensions which typically are in between these two values.
In contrast, the values deviate from the 2d Ising model values in the opposite direction as the mean
field values, indicate that fluctuations are perhaps more pronounced. This is consistent with our
interpretation of the geometry. Of course, it should be noted that our estimates are still approximate, 
primarily due to the difficulty in controlling the finite size effects. However, the key point is that
the variation of the exponent $\gamma/\bar\nu$ is within $1\sigma-2\sigma$ for all the tessellations. 
The exponent $1/\bar\nu$ shows a wider spread of up to about $3\sigma$, which points to stronger finite-size 
effects, but does not disagree with the hypothesis of the existence of a new universality class. We 
will also illustrate that significant finite-size effects show up in other observables, to a much larger 
extent than is observed for flat geometries. Moreover, such finite-size effects have also been studied
in the literature \cite{PhysRevB.44.5081, Hasenbusch_1999, Hasenbusch_2010}.

 It is usually the case that on smaller lattice sizes, there are sub-leading operators which contribute
to the scaling function. This is a well-known issue in the context of models in flat space, and 
can be addressed using scaling corrections. Given the large finite size effects in negatively curved
spaces, it is also important to check whether scaling corrections applied to data on smaller lattices
can confirm whether the extracted values of the critical exponents are robust. We applied this idea
to the case of $\{5,4\}$ tessellation which showed the largest deviation in
\cite{dey2026boundaryconditiondependentuniversality}. The correction-to-scaling term provided an
improved estimate for the critical exponents $1/\bar\nu=0.170(2)$, $\gamma/\bar{\nu} = 0.743(4)$,
and  $\beta_c=0.630(1)$, in agreement with the largest lattices, and also the exponent 
obtained from other tessellations (please see the Supp. Mat. of 
\cite{dey2026boundaryconditiondependentuniversality} for further details).
A similar analysis for the $\{3,7\}$ tessellation leads to the same conclusion. Using the three smaller lattice sizes, $N=4264$, $11173$, and $29261$, a fit without correction-to-scaling terms gives $\beta_c=0.301(1)$, $\gamma/\bar\nu=0.616(9)$, and $1/\bar\nu=0.199(1)$. Including the leading correction-to-scaling term shifts these estimates to $\beta_c=0.312(8)$, $\gamma/\bar\nu=0.740(76)$, and $1/\bar\nu=0.169(19)$, bringing them into agreement with the estimates obtained from the largest lattices, as reported in \cref{tab:fss_exponents}.
\begin{table*}
  \centering
  \setlength{\tabcolsep}{14pt}
  \renewcommand{\arraystretch}{1.4}
  \begin{tabular}{cccccc}
    \hline
    Lattice & Order($n$) & $\beta_c$ & $\gamma/\bar{\nu}$ & $1/\bar{\nu}$ & $\chi^2/\mathrm{d.o.f.}$ \\
    \hline
    \multirow{2}{*}{$\{3,7\}$} & 4th &  0.313(1) &  0.745(10)& 0.177(2) & 1.97 \\
                               & 6th &  0.313(1) &  0.747(10)& 0.177(2) & 1.97   \\
    \hline
    \multirow{2}{*}{$\{3,8\}$} & 4th &  0.315(1) &  0.73(1) &0.170(2)  &1.82  \\
                               & 6th &  0.315(1) &  0.73(1) &0.170(2)  & 1.82 \\
    \hline
    \multirow{2}{*}{$\{5,4\}$} & 4th &  0.632(3) &  0.743(12) &0.162(3)  &1.79  \\
                               & 6th &  0.631(3) &  0.741(12) &0.162(3)  & 1.81 \\
    \hline
    \multirow{2}{*}{$\{6,6\}$} & 4th & 0.723(2) & 0.738(3) &0.172(1)  & 2.34 \\
                               & 6th & 0.722(2) & 0.737(3) & 0.171(1) & 2.06 \\
    \hline
  \end{tabular}
  \caption{Finite-size scaling parameters for the $\{3,7\}$, $\{3,8\}$, $\{5,4\}$, and $\{6,6\}$ tilings,
           obtained from data collapse with 4th- and 6th-order scaling polynomials with OBC.}
  \label{tab:fss_exponents}
\end{table*}

\begin{figure}[!tbh]
    \centering
    \includegraphics[width=\linewidth]{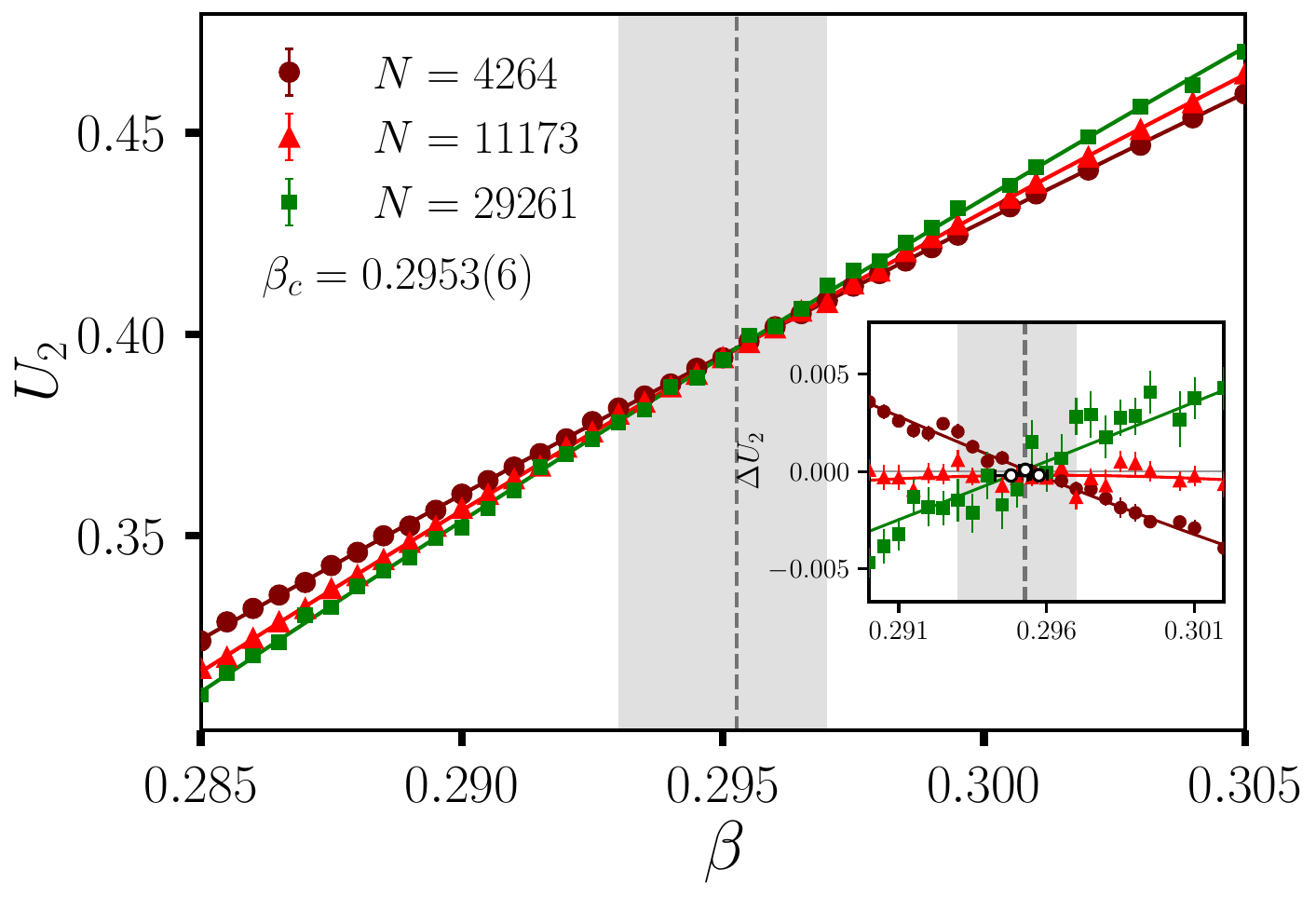}
    \includegraphics[width=\linewidth]{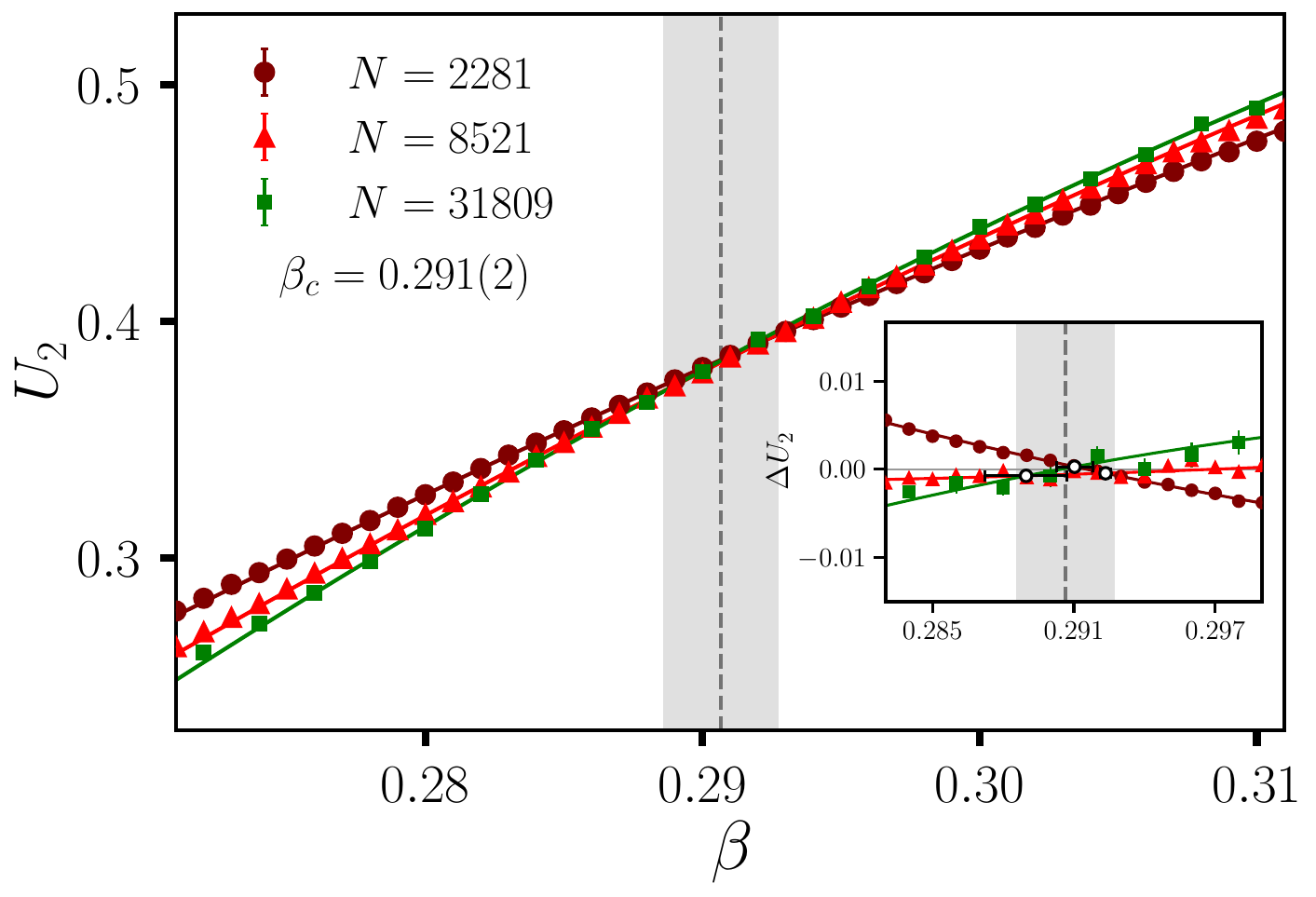}
     \includegraphics[width=\linewidth]{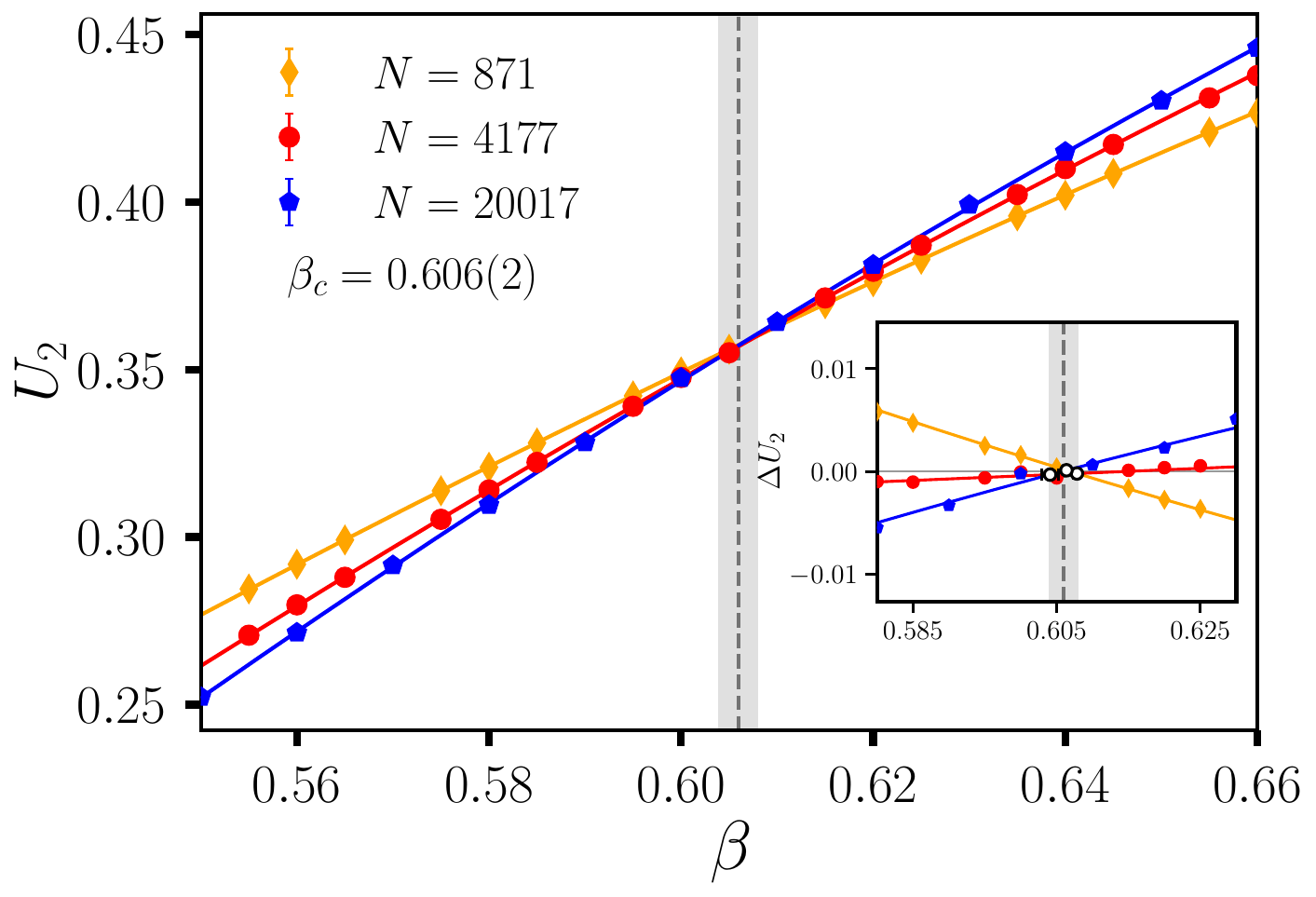}
    \caption{Crossing of Binder cumulant $U_2$ for the (top) $\{3,7\}$, (middle) 
    $\{3,8\}$ and (bottom) $\{6,6\}$ tessellation with OBC.}
    \label{fig:crossing_binder_OBC}
\end{figure}

As an independent estimate of the critical point, we have also studied the Binder cumulant $U_2$,
which is expected to become independent of the system size $N$ at the transition, so that the curves 
for different $N$ intersect. In the thermodynamic limit, on flat lattices, independent estimates
of the critical point should all agree. \cref{fig:crossing_binder_OBC} shows $U_2$ for the 
$\{3,7\}$, $\{3,8\}$ and $\{6,6\}$ tessellations; the corresponding analysis for $\{5,4\}$ was 
presented in \cite{dey2026boundaryconditiondependentuniversality}. 
The statistical error of each crossing is obtained using the bootstrap method, where the estimate 
of critical point is taken as the mean of the crossings of neighbouring lattice sizes 
\cite{1981ZPhyB..43..119B, 10.1063/1.3518900}, with the spread of these crossings quoted as the error.
The insets of each \cref{fig:crossing_binder_OBC} show $\Delta U_2 = U_2 - \langle U_2\rangle$, where
$\langle U_2\rangle$ is the average of the fitted curves, which makes the crossings directly visible.
From the crossings, we obtain the critical couplings as $\beta_c = 0.2953(6)$, $\beta_c = 0.291(2)$, and 
$\beta_c = 0.606(2)$ for the $\{3,7\}$, $\{3,8\}$ and $\{6,6\}$ tessellations respectively, and 
$\beta_c = 0.591(2)$ for $\{5,4\}$ tessellation mentioned in \cite{dey2026boundaryconditiondependentuniversality}. 
For all four tessellations, these values are significantly lower than the critical points obtained from 
the susceptibility $\chi$, tabulated in \cref{tab:fss_exponents}. 

 The most likely reason for this shift is finite-size corrections. Computing $U_2$ requires the fourth 
moment of the magnetization, which we have computed using local Metropolis update. Owing to its critical
slowing down (\cref{fig:tau_int}), the Binder analysis is restricted to smaller lattices than the FSS of 
$\chi$, for which the worm update is employed. On these smaller lattices the corrections to scaling are 
large. This is already visible in $\chi$ itself: for $\{5,4\}$, a fit to the smaller lattices without a 
correction term gives $\beta_c = 0.607(9)$, close to the Binder crossing, whereas including the leading
correction to scaling restores $\beta_c = 0.630(1)$, in agreement with the largest lattices presented
in \cref{tab:fss_exponents}. A similar behaviour is observed for the $\{3,7\}$ tessellation. Using the smaller
lattices, the FSS of $\chi$ without a correction term gives $\beta_c = 0.301(1)$, 
closer to the Binder crossing shown in top panel of \cref{fig:crossing_binder_OBC}. Including the 
leading correction to scaling shifts the estimate to $\beta_c = 0.312(8)$, with $\gamma/\bar\nu = 0.740(76)$ and
$1/\bar\nu = 0.169(19)$, 
consistent with the results from the largest lattices in \cref{tab:fss_exponents}.
The Binder 
cumulant is even more sensitive, since the higher moments of the magnetization receive large contributions 
from the fluctuating boundary spins in OBC. This explained the necessity of a scaling correction term in 
the FSS collapse of $U_2$ for $\{5,4\}$ \cite{dey2026boundaryconditiondependentuniversality}.
The estimate of Binder crossing should therefore be studied in the future using worm algorithm, and on
bigger lattices. 

\begin{figure}[!tbh]
    \centering
    \includegraphics[width=\linewidth]{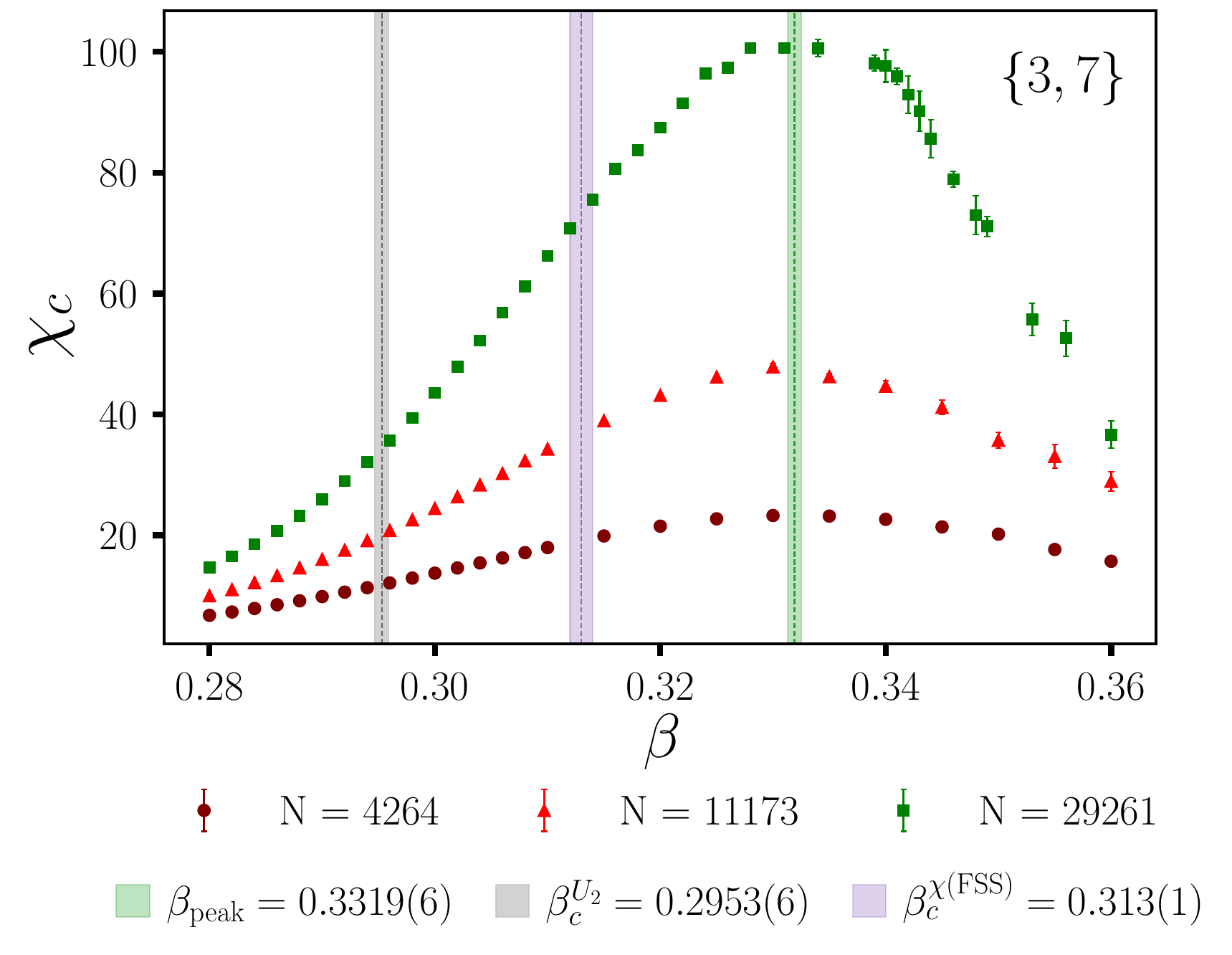}
    \caption{Connected susceptibility $\chi_c$ on $\{3,7\}$ tessellation with OBC. The peak at 
    $\beta_{\rm peak}=0.3319(6)$ from $\chi_c$ lies well above $\beta_c=0.2953(6)$ (Binder crossing \cref{fig:crossing_binder_OBC}) and $\beta_c=0.313(1)$ (FSS \cref{fig:FSS_OBC}).}
    \label{fig:connected_chi_OBC_tile37}
\end{figure}

\begin{figure}[!tbh]
    \centering
    \includegraphics[width=\linewidth]{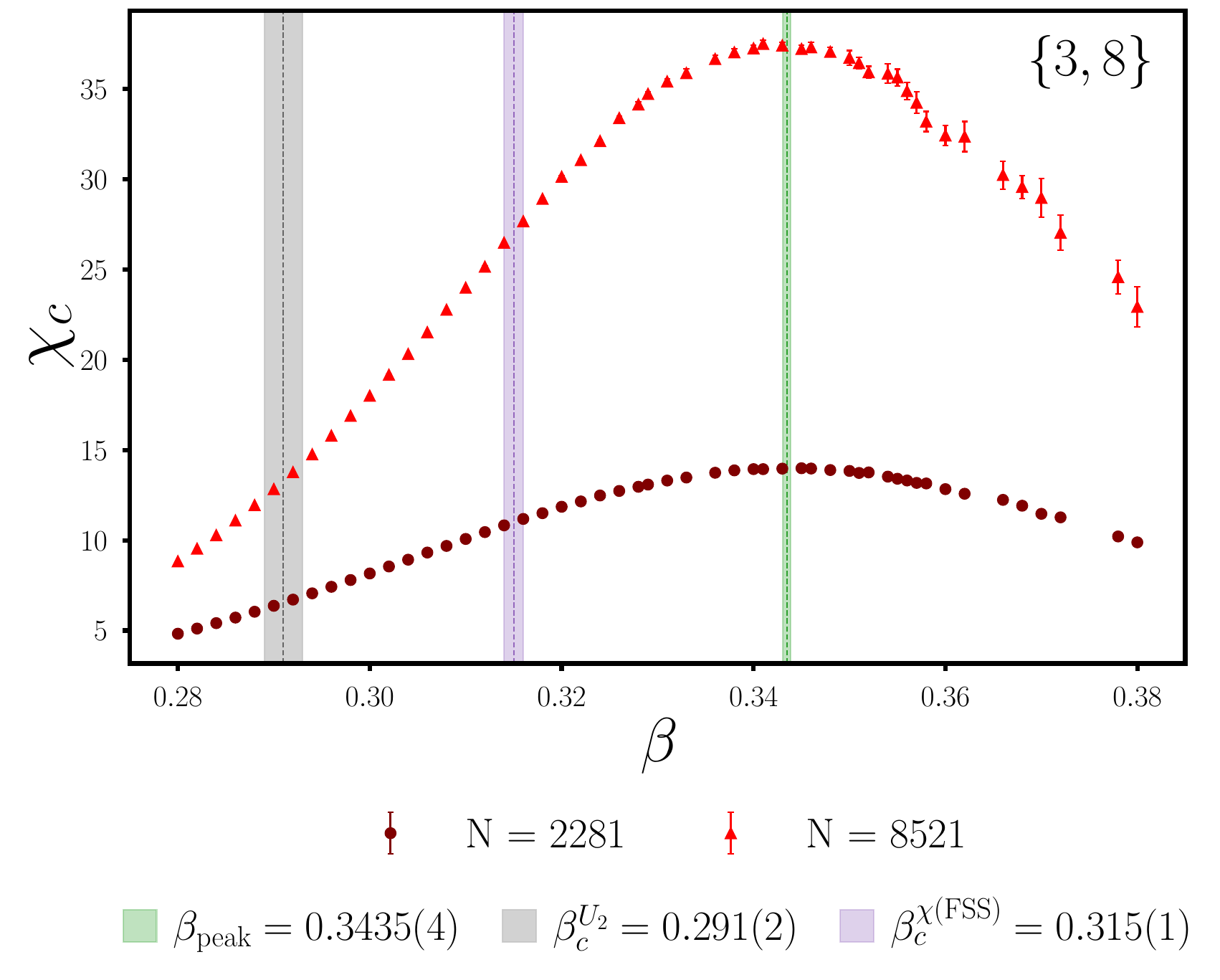}
    \caption{Connected susceptibility $\chi_c$ on $\{3,8\}$ tessellation with OBC. The peak at 
    $\beta_{\rm peak}=0.3435(4)$ from $\chi_c$ lies well above $\beta_c=0.291(2)$ (Binder crossing \cref{fig:crossing_binder_OBC}) and $\beta_c=0.315(1)$ (FSS \cref{fig:FSS_OBC}).}
    \label{fig:connected_chi_OBC_tile38}
\end{figure}

\begin{figure}[!tbh]
    \centering
    \includegraphics[width=\linewidth]{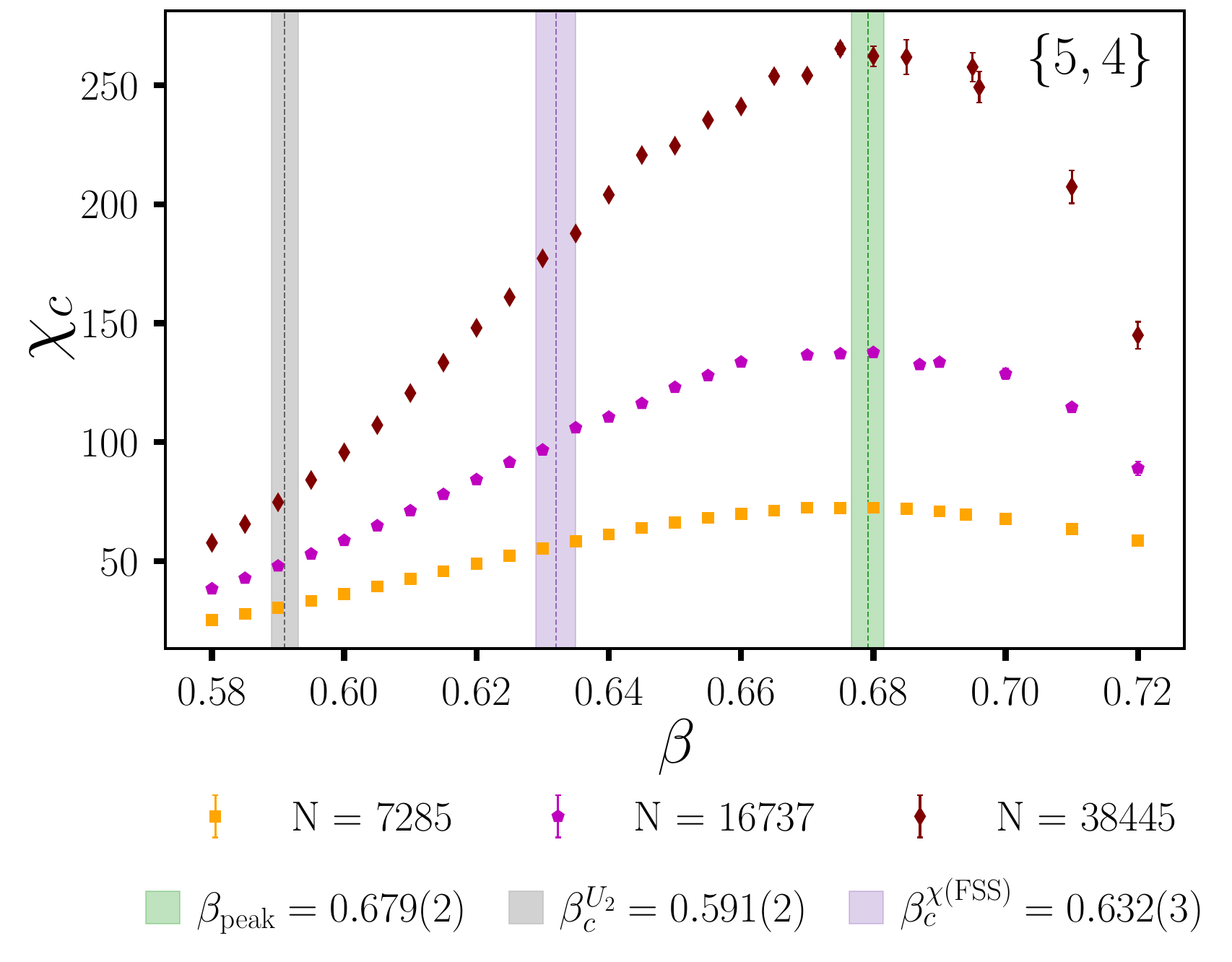}
    \caption{Connected susceptibility $\chi_c$ on $\{5,4\}$ tessellation with OBC. The 
    peak at $\beta_{\rm peak}=0.679(2)$ from $\chi_c$ lies well above $\beta_c=0.591(2)$ 
    (Binder crossing \cref{fig:crossing_binder_OBC}) and $\beta_c=0.632(3)$ (FSS \cref{fig:FSS_OBC}).}
    \label{fig:connected_chi_OBC_tile54}
\end{figure}

\begin{figure}[!tbh]
    \centering
    \includegraphics[width=\linewidth]{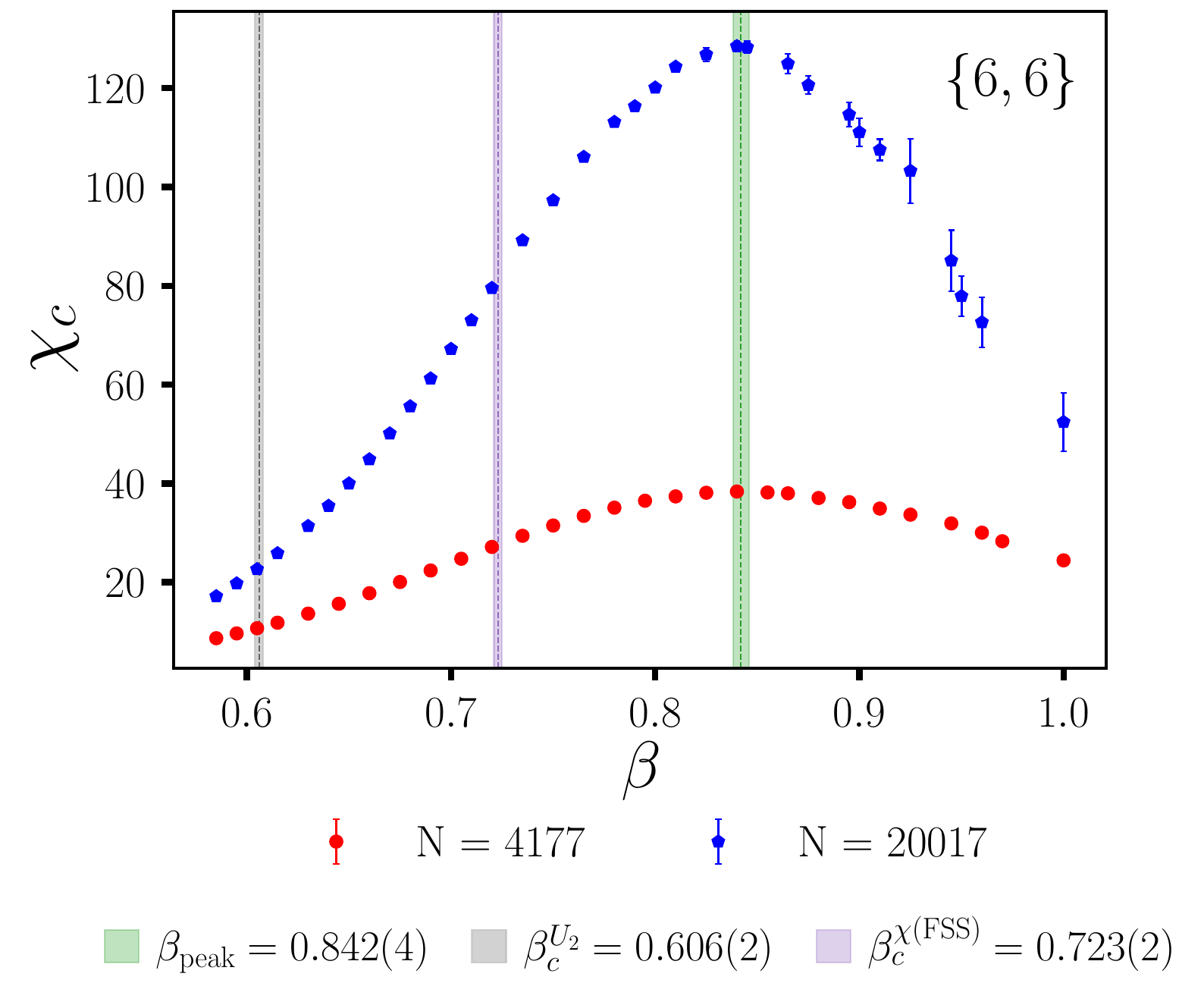}
    \caption{Connected susceptibility $\chi_c$ on $\{6,6\}$ tessellation with OBC. The peak 
    at $\beta_{\rm peak}=0.842(4)$ from $\chi_c$ lies well above $\beta_c=0.606(2)$ (Binder crossing \cref{fig:crossing_binder_OBC}) and $\beta_c=0.723(2)$ (FSS \cref{fig:FSS_OBC}).}
    \label{fig:connected_chi_OBC_tile66}
\end{figure}

 On flat space, the peak of the connected susceptibility $\chi_c$ provides an independent estimate of 
the critical coupling. Since this quantity is inaccessible to the worm algorithm, we have measured this 
using the Metropolis algorithm. Figures \ref{fig:connected_chi_OBC_tile37}, \ref{fig:connected_chi_OBC_tile38}, \ref{fig:connected_chi_OBC_tile54} 
and \ref{fig:connected_chi_OBC_tile66} show the connected susceptibility $\chi_c$ vs $\beta$ 
for the $\{3,7\}$, $\{3,8\}$, $\{5,4\}$ and $\{6,6\}$ tessellations, respectively, using the largest volumes accessible 
with local Metropolis updates. The shaded bands indicate three different estimates of critical point obtained 
from the susceptibility peak, Binder-cumulant crossing and FSS collapse of $\chi$. 
In all four tessellations, $\chi_c$ exhibits a pronounced maximum, with a gradual increase in 
the weak coupling regime $(\beta<\beta_{\rm peak})$ and a steeper decrease in the ordered regime 
$(\beta>\beta_{\rm peak})$. The position of the peak is determined for each system size and 
tessellation from a polynomial fit of $\chi_c$, yielding $\beta_{\rm peak} = 0.3319(6)$, $0.3435(4)$, $0.679(2)$ 
and $0.842(4)$  for the $\{3,7\}$, $\{3,8\}$, $\{5,4\}$ and $\{6,6\}$ tessellations, respectively. For these 
four tessellations, the critical point estimates are ordered as 
$\beta_c^{U_2} < \beta_c^{\chi({\rm FSS})} < \beta_{\rm peak}$, revealing a systematic separation 
between the three characteristic couplings. This separation suggests that the different estimators 
are affected by finite-size effects \cite{dey2026boundaryconditiondependentuniversality}.

 A similar discrepancy between the characteristic temperature scales was recently reported for 
$\{5,4\}$ tessellation with free boundary conditions in \cite{jaroszewicz2026temperaturescalesisingmodel}. 
In that study, the authors report a susceptibility-peak temperature $T_{c_2}=1.4782(15)$ and a 
higher-temperature scale $T^*\simeq 1.67(3)$ associated with Binder-cumulant crossings. In terms 
of inverse temperature, these correspond to $\beta_{c_2}\simeq 0.6765(7)$ and $\beta^{*}\simeq 0.599(11)$, 
respectively. These estimates are in remarkably good agreement with our corresponding values of 
$\beta_{\rm peak} = 0.679(2)$ and $\beta_c^{U_2} = 0.591(2)$. Thus, despite the differences in 
the numerical methods, both studies find a similar separation between the susceptibility-peak and 
Binder-crossing estimates. Our FSS estimate, $\beta_c^{\chi({\rm FSS})}=0.632(3)$, lies between 
these two values. Notably, this estimate is obtained from the largest system sizes accessible 
in our simulations, whereas the Binder-crossing and susceptibility-peak estimates are limited 
by the computational cost of local Metropolis updates and are therefore obtained from smaller 
lattices. As discussed in our studies \cite{dey2026boundaryconditiondependentuniversality}, the 
separation between $\beta_c^{U_2}$ and $\beta_c^{\chi({\rm FSS})}$ is consistent with finite-size 
corrections, since the FSS collapse of $\chi$ restricted to the smaller lattices yields an estimate 
close to the Binder crossing. The origin of the separation of $\beta_{\rm peak}$ is less clear. 
It may reflect a distinct thermodynamic scale \cite{jaroszewicz2026temperaturescalesisingmodel}, or 
it may arise from finite-size effects that vanish in the thermodynamic limit, this remains an open question.

\subsection{Universality with WBC}\label{sec:WBC}
 The use of auxiliary spin in the WBC provides a direct way to connect all boundary spins. The primary 
consequence is that the independent fluctuations of the boundary spins are suppressed, while a collective mode, 
in which the boundary fluctuates together with $s_{\rm aux}$, survives. The global magnetization is therefore 
dominated by the extensive boundary, and does not directly reflect the ordering in the bulk. A more suitable 
probe is the correlation between the central spin and the auxiliary spin, $\braket{s_0 s_{\rm aux}}$, which 
tests whether the effects of the boundary reaches the center of the lattice. As this correlation extends over 
the entire radial direction, the number of layers $L$ serves as the natural finite-size variable (instead of
$N$), and one expects, 
\begin{equation}
  \braket{s_0 s_{\rm aux}} = L^{-\beta_m/\bar\nu}\, \Phi\!\left(t\,L^{1/\bar\nu}\right) ,
  \label{eq:FSS_s0saux}
\end{equation}
where $\beta_m$ denotes the critical exponent corresponding to magnetization. For the $\{5,4\}$ tessellation, 
FSS of $\braket{s_0 s_{\rm aux}}$ yields $\beta_c=0.348(4)$, $\beta_m=0.48(7)$ and $1/\bar\nu=1.03(6)$  
\cite{dey2026boundaryconditiondependentuniversality}. These exponents are consistent with the mean-field values.
We checked the consistency of this fit by fixing $\beta_m$ to its mean-field value, and redoing the fit, which
yielded $\beta_c=0.347(1)$ and $1/\bar\nu=1.05(2)$ \cite{dey2026boundaryconditiondependentuniversality}. 

\begin{figure}[!tbh]
    \centering
    \includegraphics[width=\linewidth]{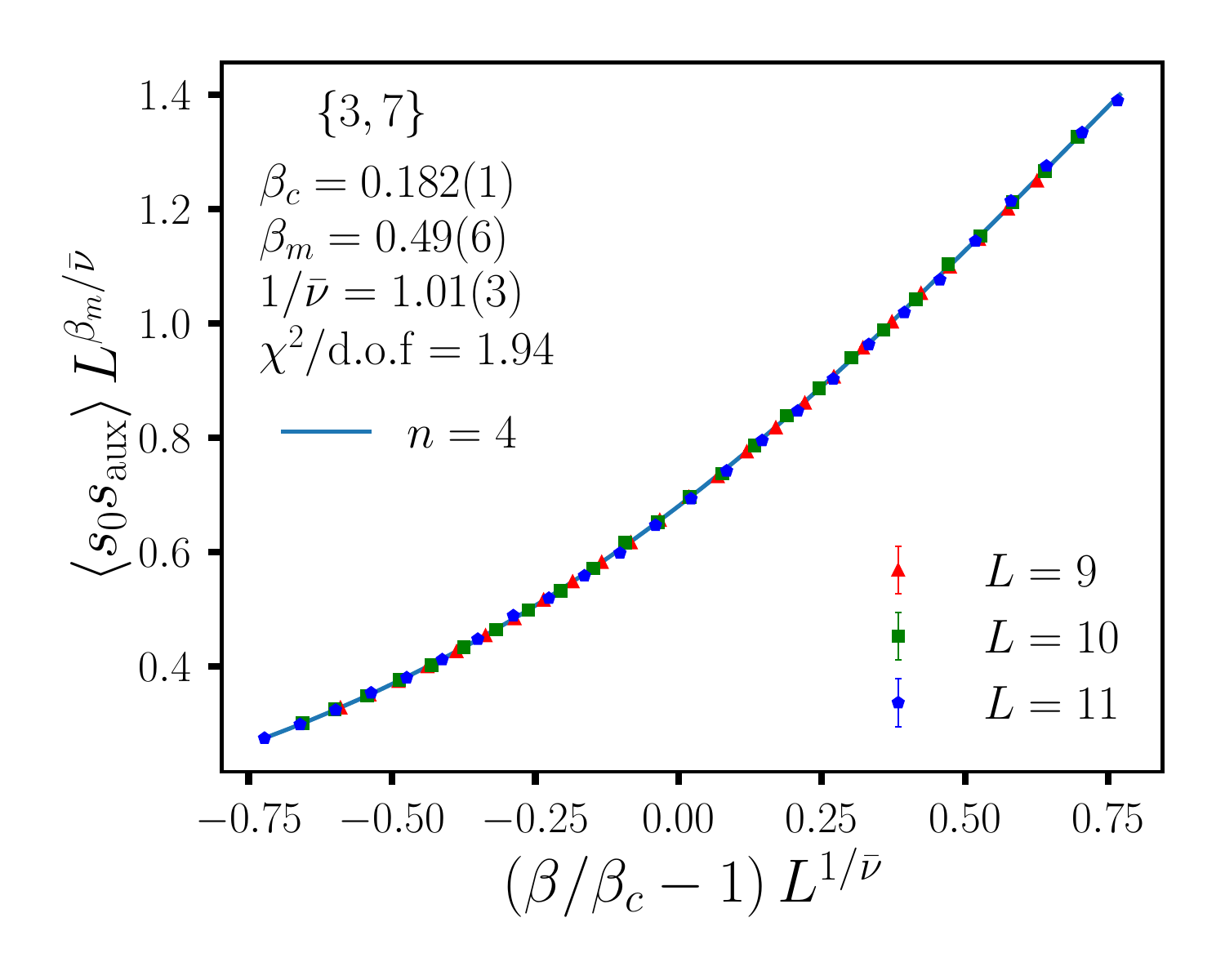}
    \includegraphics[width=\linewidth]{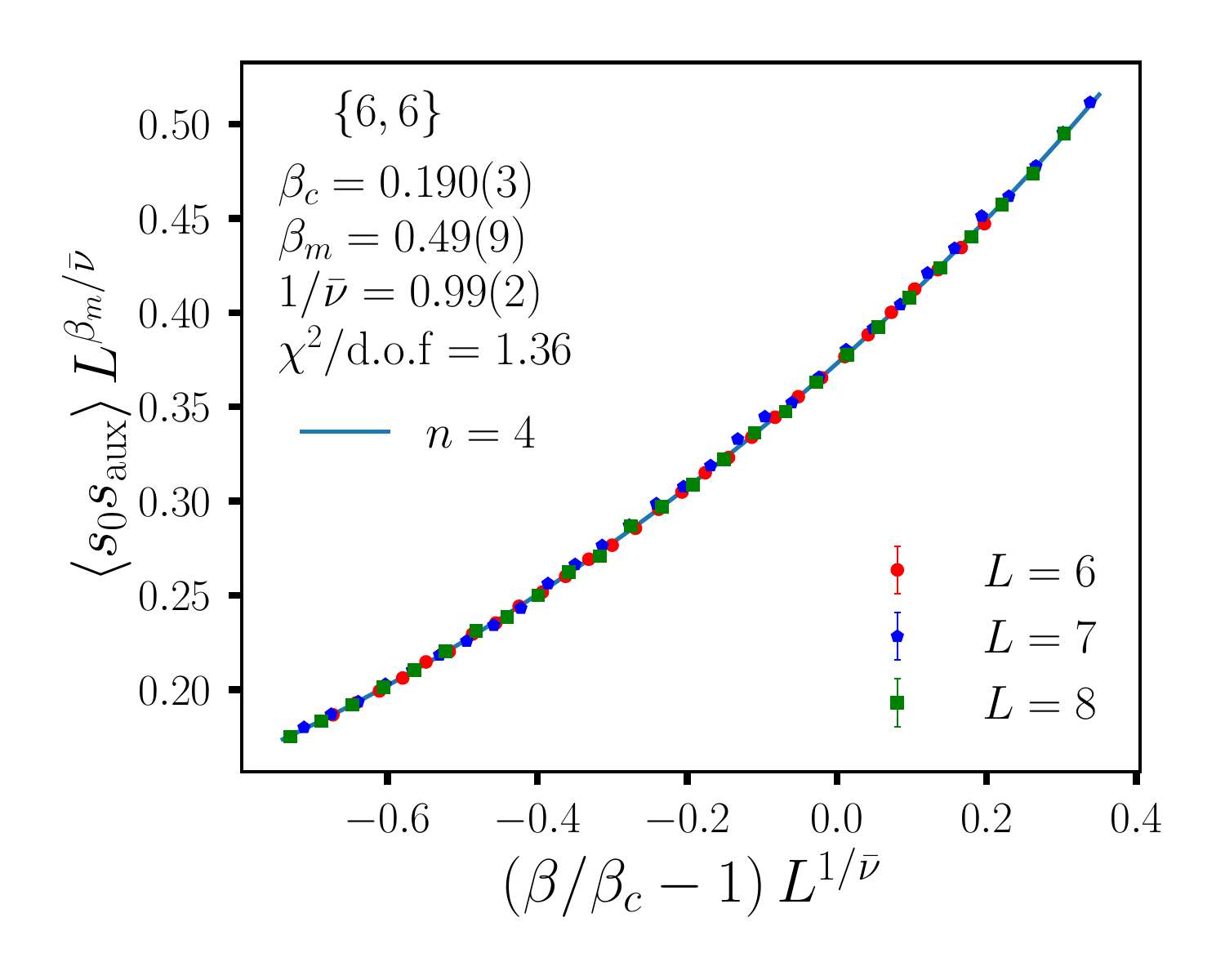}
    \caption{FSS of the two-point correlation function $\braket{s_0 s_{\rm aux}}$ of the spins separated by the 
    largest geodesic distance for the $\{3,7\}$ (top) and $\{6,6\}$ (bottom) tessellations, yields critical 
    exponents consistent with mean-field scaling.}
    \label{fig:FSS_corr0_WBC}
\end{figure}

 We repeated the analysis for the $\{3,7\}$ and $\{6,6\}$ tessellations, depicted in \cref{fig:FSS_corr0_WBC}
in order to observe the robustness against different tessellations. For the $\{3,7\}$ tessellation with $L=9$-$11$, 
the collapse yields $\beta_c=0.182(1)$, $\beta_m=0.49(6)$ and $1/\bar\nu=1.01(3)$. On the other hand, for the 
$\{6,6\}$ tessellation, we obtain $\beta_c = 0.190(3) $, $\beta_m = 0.49(9)$ and $1/\bar \nu = 0.99(2)$.
In both tessellations, the exponents agree with the mean-field values $\beta_m=1/2$ and $1/\bar\nu=1$ within errors. 
In \cref{tab:betac_cayley_WBC} we compare critical point $\beta_c^{\rm WBC}$ of $\{p,q\}$ tessellations with WBC, with the q-Cayley-tree critical point
$\beta_c^{\rm tree}$ (\cref{eq:beta_c1_tree}, discussed in \cref{sec:eggarter}). The two values agree within $1\sigma$ for the $\{5,4\}$ tessellation, whereas they differ by approximately $2$--$4\sigma$ for $\{3,8\}$ and $\{6,6\}$ tessellations, and by a larger amount for $\{3,7\}$ tessellation. Nevertheless, the overall proximity of $\beta_c^{\rm WBC}$ and $\beta_c^{\rm tree}$ suggest that the Cayley-tree threshold provides a useful reference for the WBC transition on finite hyperbolic tessellations. The remaining discrepancies may reflect the effects of closed loops, which are absent on the Cayley tree, as well as finite-size effects.

\begin{figure}[!tbh]
    \centering
    \includegraphics[width=\linewidth]{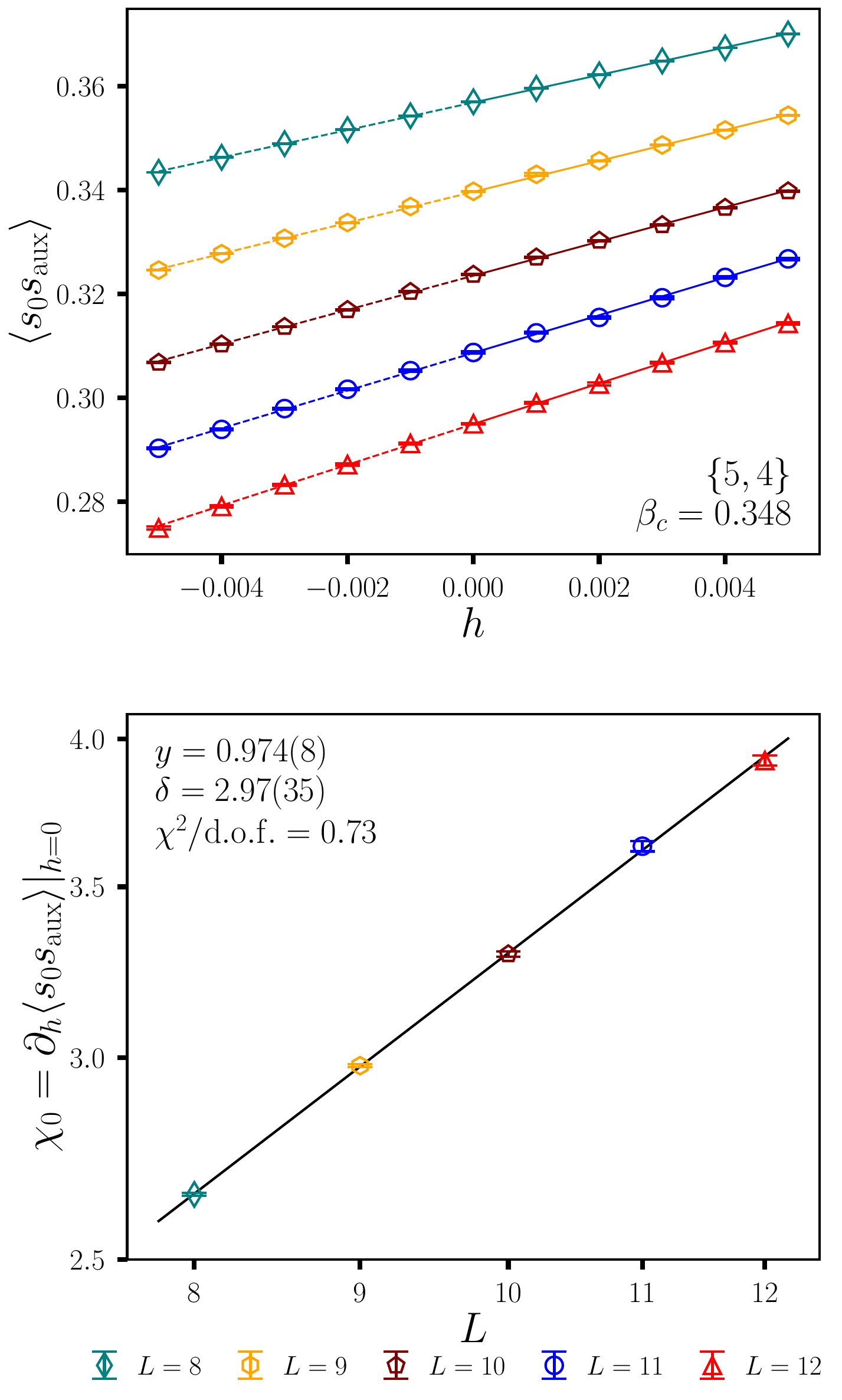}
    \caption{Response of the central spin to a uniform field $h$ with WBC on the
    $\{5,4\}$ tessellation at $\beta_c=0.348$. Top: $\langle s_0 s_{\rm aux}\rangle$ depends linearly
    on $h$ for each $L$, and the slope defines $\chi_0(L)$. Bottom: $\chi_0$ grows as
    $L^{y}$, which yields $\delta=2.97(35)$, close to the mean-field value $\delta=3$. 
    We obtain the statistical errors ($\mathcal{O}(10^{-4})$ for $\langle s_0 s_{\rm aux}\rangle$,
$\lesssim 0.5\%$ for $\chi_0$) are smaller than the symbols.}
    \label{fig:delta_WBC}
\end{figure}

\begin{table}[!tbh]
    \centering
    \renewcommand{\arraystretch}{1.0}
    \begin{tabular}{c c c}
    \hline
     Lattice    & $\beta_c^{\rm tree}$ & $\beta_c^{\rm {WBC}}$ \\
     \hline
     $\{3,7\}$  & 0.168236 & 0.182(1) \\
     $\{3,8\}$  & 0.14384 &  0.146(1)\\
     $\{5,4\}$ & 0.34657 & 0.348(4)\\
     $\{6,6\}$ & 0.2027 & 0.190(3) \\
     \hline
    \end{tabular}
    \caption{Comparison of critical point $\beta_c^{\rm WBC}$ obtained from WBC on $\{p,q\}$ tessellations with the corresponding critical values $\beta_c^{\rm tree}$ (\cref{eq:beta_c1_tree}) for the $q$-Cayley tree.}
    \label{tab:betac_cayley_WBC}
\end{table}
 The mean-field character of the transition can be examined further through the critical 
isotherm exponent $\delta$. To define the appropriate field response, we distinguish the 
full WBC ensemble from the ensemble where $s_{\rm aux}=+1$. In the latter, the central-to-boundary 
correlator equals the central magnetization,
\begin{equation}
   m_{0,+}(h,L) \equiv \braket{s_0}_{h,s_{\rm aux}=+1} = \braket{s_0s_{\rm aux}}_{h,s_{\rm aux}=+1}.
  \label{eq:m0_sector}
\end{equation}
We apply a uniform field $h$ to the physical spins. Within the positive auxiliary-spin sector, 
the Hamiltonian is
\begin{equation}
  H_+(h) = H_{\rm OBC} - J\sum_{i\in\partial V_N}s_i - h\sum_{i=1}^{N}s_i.
  \label{eq:H_wbc_h}
\end{equation}
Thus, the auxiliary coupling acts as a fixed positive boundary field, while $h$ probes the response 
to a uniform field throughout the physical lattice.

This distinction between ensembles is essential. In the complete dynamical-auxiliary ensemble, the 
simultaneous reversal of all spins maps $h$ to $-h$ while leaving $s_0s_{\rm aux}$ invariant.
Consequently, the full correlator is even in $h$ and has no linear response at $h=0$ on a 
finite lattice. In the conditional ensemble, however, the positive boundary field breaks 
this symmetry, and a linear response of $m_{0,+}$ is allowed. At zero field, symmetry relates 
the conditional magnetization to the full WBC correlator,
\begin{equation}
  m_{0,+}(0,L) = \braket{ s_0s_{\rm aux}}_{h=0}.
  \label{eq:m0_zero_field}
\end{equation}
This equality connects the present analysis to the zero-field scaling studied above. 

 At the critical coupling, the critical isotherm exponent is defined through the response,
\begin{equation}
  m_{0,+}^{(\infty)}(h) \equiv \lim_{L\to\infty}m_{0,+}(h,L) \sim h^{1/\delta},~~h\to0^+,~~\beta=\beta_c.
  \label{eq:critical_isotherm}
\end{equation}
Here the thermodynamic limit is taken before the field is sent to zero. To connect this behaviour 
to finite lattices, we test the radial finite-size scaling hypothesis
\begin{equation}
  m_{0,+}(h,L) = L^{-a}\Psi_+\!\left(hL^{a\delta}\right),~~~a=\frac{\beta_m}{\bar\nu},
  \label{eq:FSS_h}
\end{equation}
up to corrections to scaling. The large-positive-argument behaviour $\Psi_+(x)\sim x^{1/\delta}$ reproduces
\cref{eq:critical_isotherm}: taking $L\to\infty$ at fixed $h>0$ sends $x=hL^{a\delta}$ to infinity and cancels 
the prefactor $L^{-a}$, yielding $m_{0,+}^{(\infty)}(h)\sim h^{1/\delta}$. By contrast, the finite-size linear 
response follows from the small-$x$ expansion, obtained by taking $h\to0$ at fixed $L$,
\begin{equation}
  m_{0,+}(h,L) = L^{-a}\left[\Psi_+(0) + \Psi_+'(0)hL^{a\delta} + \mathcal{O}\!\left((hL^{a\delta})^2\right) \right].
  \label{eq:Psi_expansion}
\end{equation}
The nonzero intercept is the boundary-induced central magnetization at finite $L$, whereas the slope defines the
conditional response
\begin{equation}
  \chi_{0,+}(L) \equiv \left.\frac{\partial m_{0,+}(h,L)}{\partial h}\right|_{h=0} = \Psi_+'(0)L^y,~~
  y=a(\delta-1).
  \label{eq:chi0_scaling}
\end{equation}
Provided the leading linear coefficient is nonzero, the scaling
of this response therefore yields
\begin{equation}
  \delta=1+\frac{y}{\beta_m/\bar{\nu}}.
  \label{eq:delta_from_y}
\end{equation}
This provides a test of the mean-field value $\delta=3$ using the scaling hypothesis based on the externally
applied field. 

In \cref{fig:delta_WBC}, we present the response of the central-to-boundary correlator obtained by 
simulating \cref{eq:H_wbc_h} with the local Metropolis update at $\beta=0.348$, for $L=8-12$ with 
fields $-0.005\le h\le 0.005$ in steps of $0.001$. Because $s_{\rm aux}$ couples to all boundary
spins, a single-spin update of $s_{\rm aux }$ costs an energy extensive in the boundary and is never 
accepted, even at $h=0$. The measurements therefore probe the positive auxiliary-spin sector rather 
than the unrestricted finite-system ensemble. Positive and negative values of $h$ respectively reinforce 
and oppose the positive boundary bias. We extract $\chi_{0,+}(L)$ from the regime around $h=0$
and determine its size dependence. 

 As shown in \cref{fig:delta_WBC} (top), $\braket{s_0 s_{\rm{aux}}}$ is linear in $h$ for all $L$, 
confirming that \cref{eq:Psi_expansion} applies. The slope $\chi_0$ grows as $L^{y}$, as predicted by 
\cref{eq:chi0_scaling}, with $y=0.974(8)$ and $\chi^2/\mathrm{d.o.f.}=0.73$ (bottom). Using 
$\beta_m =0.48(7)$ and $1/\bar \nu = 1.03(6)$ from the collapse of $\braket{s_0 s_{\rm aux}}$ 
with \cref{eq:FSS_s0saux} \cite{dey2026boundaryconditiondependentuniversality}, 
\cref{eq:delta_from_y} yields $\delta=2.97(35)$, consistent with the mean-field value $\delta=3$.
Together with the exponent $\beta_m$ and $1/\bar \nu $, the critical exponent $\delta$ therefore 
confirms that the WBC phase transition belongs to the mean-field universality class.

\subsection{Effect of h: Eggarter phase}\label{sec:eggarter}

\begin{figure*}[!tbh]
    \centering
    \includegraphics[width=\linewidth]{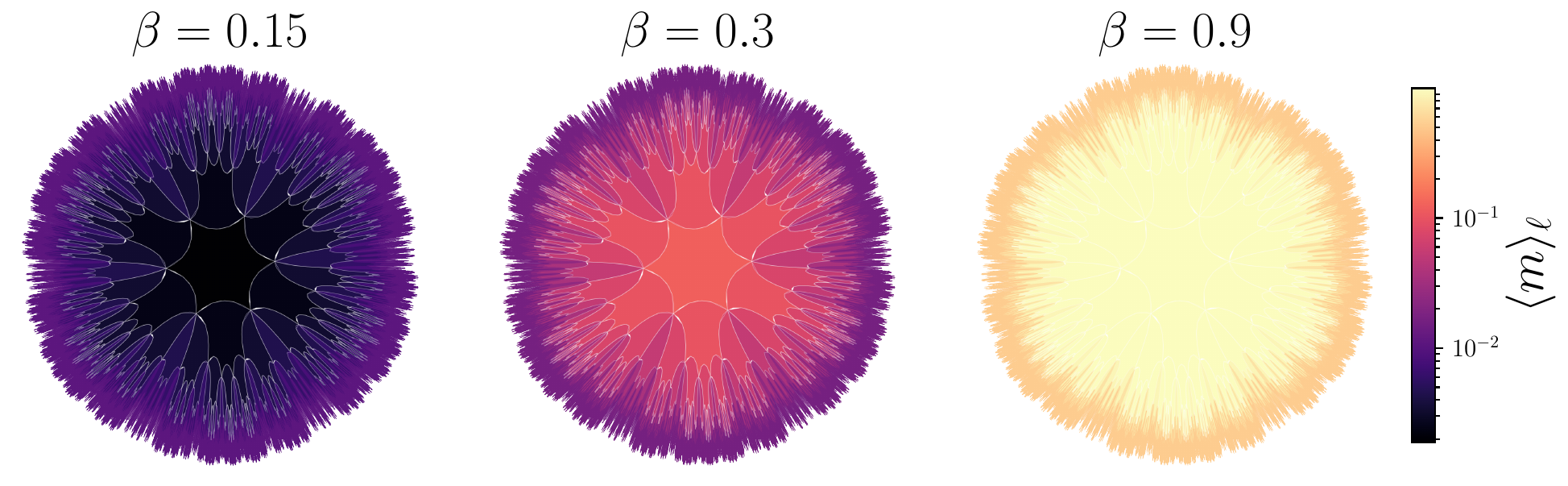}
    \caption{Qualitative evidence for the intermediate Eggarter phase. Mean magnetization per layer, 
    $\braket{m}_\ell$, on the $\{6,6\}$ tessellation ($L=7$) for $h_\partial=0.01$. At $\beta=0.15$ 
    (left), the magnetization is largest at the boundary and decays towards the inner layers. At 
    $\beta=0.3$ (middle), the central layers are more strongly magnetized than the boundary, as expected 
    in the Eggarter phase. At $\beta=0.9$ (right), the whole lattice is ordered.}
    \label{fig:heatmap_egg}
\end{figure*}

To investigate how the boundary influences the interior, we follow an alternative approach 
in which a weak magnetic field $h_\partial$ is applied to the boundary sites with OBC, as 
described in \cref{eq:H_mag}. Since the bulk sites are not subject to the field directly, 
their response is mediated through the nearest-neighbour interactions. We are particularly 
interested in characterizing how the magnetization induced by the boundary field propagates 
from the boundary towards the centre. This question was first addressed by T. P. Eggarter 
for the Ising model on the Cayley tree \cite{Eggarter1974}, and more recently by Wang
\emph{et al.} for hyperbolic lattices~\cite{f4sj-rwvj}, leading to the identification of 
the so-called Eggarter phase discussed below. 

 The Cayley-tree argument provides a simple explanation of how a local boundary field can 
induce ordering in the interior. A Cayley tree with coordination number $q$ is a connected 
graph without closed loops, in which each interior site has $q$ neighbours and each boundary 
site has a single neighbour. Geometrically, it can be viewed as the limiting case of a $\{p,q\}$ 
tessellation as $p\to\infty$. The tree is constructed outward from a central site, which forms
layer $\ell=0$. The central site is connected to $q$ sites that form the first layer $\ell = 1$.
Each site of layer $\ell\geq1$ is connected to one site of inner layer $\ell-1$ and to $q-1$ 
sites of layer $\ell+1$, so the layer $\ell\geq1$ therefore contains $N_\ell=q(q-1)^{\ell-1}$ 
sites. For a finite tree consisting of $L$ layers, $\ell=0, 1, \dots,L-1$, therefore, it has 
its boundary at $\ell=L-1$. 

 On the Cayley tree, the free energy remains analytic at all temperatures, yet the interior 
orders in the presence of an infinitesimal boundary field \cite{Eggarter1974}. Each boundary 
site at $L-1$ is connected to the central site by a path of $L-1$ bonds. Along any such path, 
the spin-spin correlation decays as $(\tanh\beta)^{L-1}$, whereas the number of boundary sites 
grows as $(q-1)^{L-1}$. To linear order in $h_\partial$, the total influence of the boundary 
on the central spin is governed by the factor $\left[(q-1)\tanh\beta\right]^{L-1}$. Since a 
boundary spin couples directly to the field, with a leading-order response
$\braket{s}_{L-1}\simeq\beta h_\partial$, the magnetization of the central spin relative to
that of a boundary spin is
\begin{equation}
  \frac{\braket{s}_0}{\braket{s}_{L-1}} \simeq N_{L-1}(\tanh\beta)^{L-1}
  = \frac{q}{q-1}\left[(q-1)\tanh\beta\right]^{L-1}.
  \label{eq:tree_ratio}
\end{equation}

For $(q-1)\tanh\beta < 1$, the ratio in \cref{eq:tree_ratio} vanishes as $L\to\infty$ and the effect 
of boundary field decays towards central layer, while for $(q-1) \tanh\beta > 1$, the
ratio grows with $L$, so the magnetization increases towards the centre and the interior
becomes magnetized along the boundary field.  The two regimes are separated by,
\begin{equation}
  \beta_{c}^{\rm tree} = \tanh^{-1}\frac{1}{q-1}.
  \label{eq:beta_c1_tree}
\end{equation}
as obtained by Eggarter \cite{Eggarter1974}.

 The Cayley tree picture extends to hyperbolic lattices, where the extensive boundary allows 
boundary effects to persist in the thermodynamic limit. As discussed by Wang \emph{et al.} 
\cite{f4sj-rwvj}, the phase structure of the Ising model on hyperbolic lattices under OBC, 
in the presence of an infinitesimal magnetic field applied to the boundary spins, exhibits 
three distinct regimes rather than the conventional distinction between paramagnetic and 
ferromagnetic phases alone. For $\beta<\beta_{c1}$, the bulk remains paramagnetic, and the 
influence of the boundary field vanishes towards the centre. For $\beta_{c1}<\beta<\beta_{c2}$, 
the system exhibits an intermediate, boundary-sensitive regime known as the Eggarter phase, 
in which the boundary field induces magnetic order in the bulk despite the absence of spontaneous 
ferromagnetic order. This is analogous to the boundary-induced ordering on the Cayley tree for 
$\beta>\beta_c^{\rm tree}$. Finally, for $\beta>\beta_{c2}$, the bulk develops spontaneous 
ferromagnetic order, which persists even in the absence of a boundary field.
For the $\{6,6\}$ tessellation, Wang \emph{et al.} \cite{f4sj-rwvj} located these two transitions 
at $\beta_{c1}=0.2029$ and $\beta_{c2}=0.8044$ in the thermodynamic limit, using a 
symmetry-restricted CTMRG method to probe observables deep in bulk. 

 The results of Wang \emph{et al.} \cite{f4sj-rwvj} establish the existence of the Eggarter 
phase in the thermodynamic limit. Here, we investigate how the three regimes manifest themselves 
on finite hyperbolic lattices by examining the spatial profile of the magnetization induced by 
a weak boundary field. In particular, we focus on how the boundary response propagates towards 
the centre and how this response changes with the strength of the boundary field. This allows 
us to distinguish boundary-induced order from spontaneous ferromagnetic order and to examine 
the characteristic spatial structure of the Eggarter phase. 

We consider finite $\{6,6\}$ tessellation with $L=7$ layers and measure the mean magnetization per layer,
\begin{equation}
  \braket{m}_\ell = \frac{1}{N_\ell}\sum_{i\in\ell}\braket{s_i},
  \label{eq:layer_mag}
\end{equation}
where $N_\ell$ denotes the number of sites in layer $\ell$, with $\ell=0,\dots,L-1$. Here $\ell=0$ 
is the central polygon (\cref{sec:geometry}) and $\ell=L-1$ is the boundary layer with $\partial V_N$ 
boundary sites. We examine three representative couplings, $\beta=0.15$, $0.3$, and $0.9$, 
corresponding to the paramagnetic, Eggarter and ferromagnetic phases, respectively.

\begin{figure}[!tbh]
    \centering
    \includegraphics[width=\linewidth]{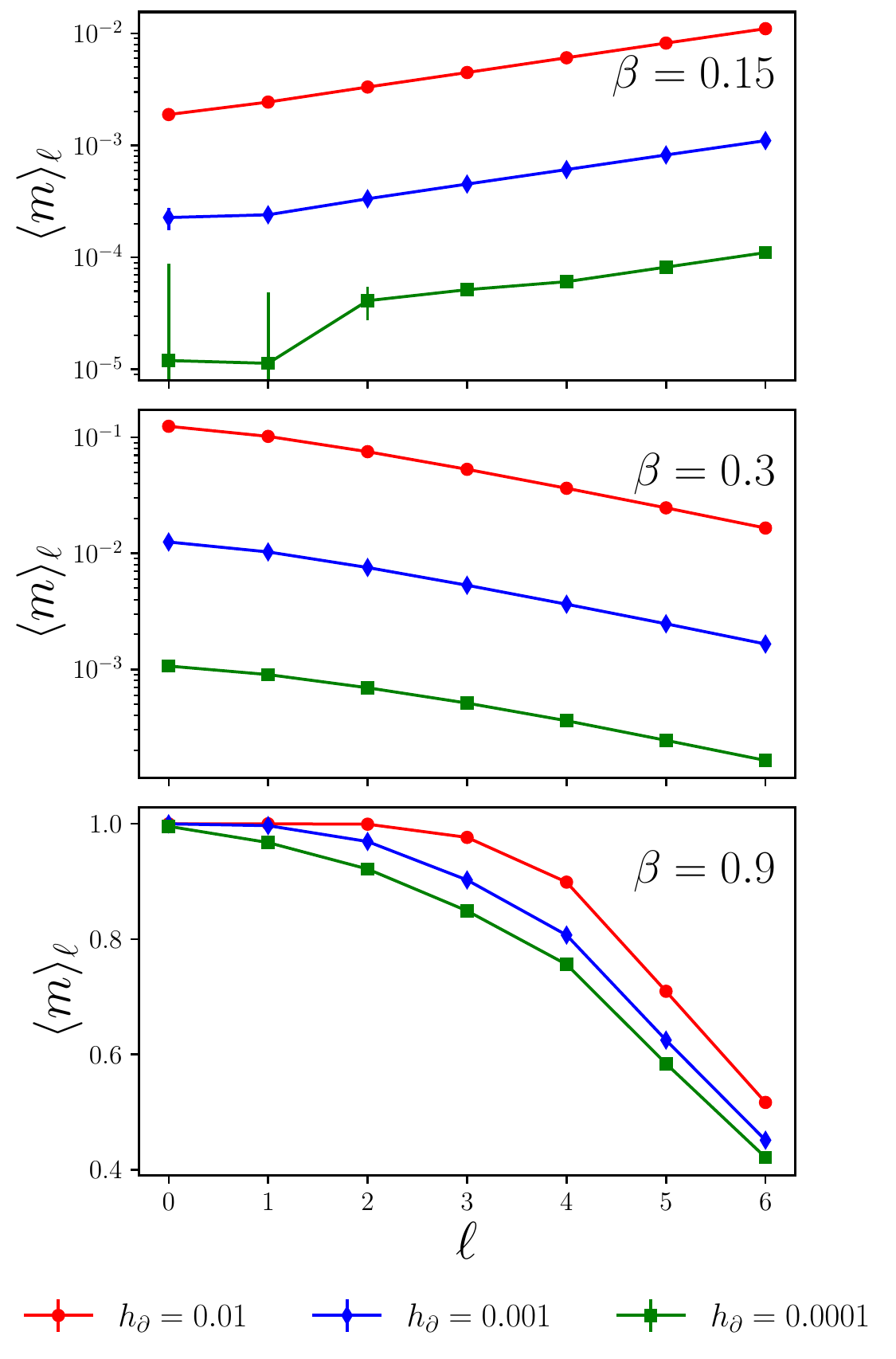}
    \caption{Mean magnetization per layer, $\braket{m}_\ell$, on the $\{6,6\}$ tessellation ($L=7$) 
    for $h_\partial=0.01$, $0.001$ and $0.0001$. For each $\beta$, the shape of the profile is nearly 
    independent of the boundary field. At $\beta=0.15$ and $\beta=0.3$, reducing $h_\partial$ by a 
    factor of ten reduces $\braket{m}_\ell$ by the same factor, whereas at $\beta=0.9$ the magnetization 
    is almost unchanged.}
    \label{fig:mag_L_tile66_egg}
\end{figure}
 The resulting magnetization profiles are shown in \cref{fig:heatmap_egg} for $h_\partial=0.01$. 
They reveal three qualitatively different responses to the boundary field. At $\beta=0.15$, the 
magnetization is largest at the boundary and decreases towards the central layer, indicating that 
the boundary influence fades as it moves inward. At $\beta=0.3$, the profile exhibits the opposite 
behaviour: the magnetization increases toward the centre with the central layer becoming strongly 
magnetized relative to the boundary itself. It provides a characteristic signature of the Eggarter phase. 
At $\beta=0.9$, finally, the entire lattice is ordered, and the magnetization is only somewhat 
reduced near the boundary, where the sites have fewer than $q$ neighbours.

 The dependence on $h_\partial$, shown in \cref{fig:mag_L_tile66_egg}, distinguishes the
induced order from the spontaneous one. At $\beta=0.15$ and $0.3$, the shape of the
profile remains unchanged as $h_\partial$ is varied from $10^{-2}$ to $10^{-4}$, while
its magnitude decreases approximately in proportion to $h_\partial$. Consequently, the magnetization
is a linear response to the boundary field in both $\beta = 0.15, 0.3$, where in the latter scenario 
the interior responds more strongly than the boundary itself. These results therefore reflect an 
enhancement of boundary-induced response, rather than spontaneous bulk ordering.
At $\beta=0.9$, on the other hand, $\braket{m}_\ell$ is almost independent of $h_\partial$, 
indicating that the lattice is spontaneously ordered.  
This persistence of bulk magnetization signals spontaneous ferromagnetic ordering, in contrast 
to the boundary-induced order observed at $ \beta=0.3$, which lies between $\beta_{c_1}$ and $\beta_{c_2}$.

\section{Conclusions and Outlook}\label{sec:conclusion}
 We have investigated how boundary conditions affect critical behaviour in the ferromagnetic 
Ising model on regular hyperbolic tessellations. The extensive boundary makes it essential to 
distinguish observables of the full lattice from those probing its interior. Our comparison 
of open and wired boundaries shows that these observables exhibit markedly different critical 
scaling, even when the physical lattice and its local interactions are unchanged.

 With open boundaries, the susceptibility of the full lattice admits a finite-size scaling 
collapse in terms of the total number of spins $N$. The extracted scaling powers differ 
substantially from mean-field volume scaling and are approximately stable across the four 
tessellations $\{3,7\}$, $\{3,8\}$, $\{5,4\}$ and $\{6,6\}$. This recurrence, despite changes 
in curvature and local connectivity, supports the possibility of a common non-mean-field 
universality class. Finite-size corrections nevertheless remain important, much more so
than on flat geometry. In particular, the different estimates of the transition coupling 
obtained from susceptibility scaling, Binder crossings and susceptibility peaks have not 
yet been shown to converge to a common thermodynamic limit. The correction-to-scaling analysis
supports the stability of the susceptibility estimates, but a controlled extrapolation of 
the other observables remains necessary.

 With wired boundaries, the extensive boundary contribution makes the global magnetization 
a less direct probe of interior ordering. The central-to-auxiliary spin correlator instead 
exhibits radial finite-size scaling consistent with a magnetization exponent $\beta_m=1/2$ on the
tessellations studied. The central-spin response to a uniform field, measured in the positive 
auxiliary-spin sector, provides a further consistency test: under the proposed field-scaling hypothesis,
its size dependence yields a value of $\delta$, compatible with the mean-field value. These 
complementary observations support a mean-field description of the WBC transition, while 
distinguishing the measured finite-size response from a direct determination of the nonlinear
critical isotherm.

 The layer-resolved response to a weak boundary field provides a further view of the 
distinction between boundary and interior behaviour. At intermediate coupling, the induced 
magnetization is amplified towards the centre, whereas at higher temperature it decreases
inwards. These profiles are qualitatively consistent with the boundary-sensitive intermediate 
regime discussed in the literature. Establishing its thermodynamic phase boundaries requires 
a joint analysis of increasing system size and decreasing boundary field.

The worm algorithm makes the large-volume susceptibility analysis practical by substantially 
reducing the critical slowing down observed with local Metropolis updates. Extending efficient 
estimators to higher magnetization moments would allow a more stringent comparison of independent 
transition diagnostics. Larger lattices, systematic correction-to-scaling studies, and a direct 
critical-isotherm measurement would further test the proposed universality classes.

 Taken together, our results show that the boundary cannot be treated merely as a finite-size 
correction on these non-amenable lattices. A characterization of their critical behaviour must 
specify both the boundary condition defining the thermodynamic limit and the observables
used to probe it.

\section{Acknowledgments}

D.B. would like to thank Simon Catterall, Alessandro Mariani, and Tim Morris for discussions 
relating to field theories on curved space-time, acknowledge STFC funded Particle Physics 
Theory Consolidated Grant ST/C004240/1 and continued support from the Alexander von Humboldt 
Foundation (Germany) in the context of the research fellowship for experienced researchers.
We thank computing resources of SINP to derive these results.
This work used the DiRAC Memory Intensive service (Cosma8/Cosma7) at Durham facility managed 
by the Institute for Computational Cosmology on behalf of the STFC DiRAC HPC Facility as well. The 
equipment was funded by BEIS capital funding via STFC capital grants ST/P002293/1, ST/R002371/1 and 
ST/S002502/1, Durham University and STFC operations grant ST/R000832/1. This work also perfomed using 
the DiRAC Data Intensive service (CSD3) at the University of Cambridge, managed by the University of 
Cambridge University Information Services on behalf of the STFC DiRAC HPC Facility (www.dirac.ac.uk). 
The DiRAC component of CSD3 at Cambridge was funded by BEIS, UKRI and STFC capital funding and STFC 
operations grants. DiRAC is part of the UKRI Digital Research Infrastructure. A.K. acknowledges the support of
the Humboldt Research Fellowship for Experienced Re-
searchers by the Alexander von Humboldt Foundation
and for the hospitality of Theoretical Physics III, De-
partment of Physics and Astronomy, Julius-Maximilians-
Universit\"{a}t W\"{u}rzburg and the support from the ICTP
through the Associates Programme (2024-2030) during
the course of this work. RS acknowledges the 
support of the Royal Society-Newton International Fellowship NIF/R1/221054-Royal Society 
during the initial stages of the project. The affiliation for RS is listed solely for 
identification. The views expressed are entirely those of the authors and do not reflect 
the official position or strategic roadmap of QuNu Labs Pvt. Ltd. This work was conducted 
independently of RS’s employment, utilising no corporate funding, resources, or proprietary 
technology from QuNu Labs Pvt. Ltd. and the latter holds no intellectual property rights 
over this manuscript. 
We acknowledge the use of LLM models such as ChatGPT 5.6 and Claude Opus 5.5
to refine the academic tone and grammar of this paper.
 

\bibliography{ref}

@article{Breuckmann_2020,
   title={Critical properties of the Ising model in hyperbolic space},
   volume={101},
   ISSN={2470-0053},
   url={http://dx.doi.org/10.1103/PhysRevE.101.022124},
   DOI={10.1103/physreve.101.022124},
   number={2},
   journal={Physical Review E},
   publisher={American Physical Society (APS)},
   author={Breuckmann, Nikolas P. and Placke, Benedikt and Roy, Ananda},
   year={2020},
   month=Feb }

@article{Prokof_ev_2001,
   title={Worm Algorithms for Classical Statistical Models},
   volume={87},
   ISSN={1079-7114},
   url={http://dx.doi.org/10.1103/PhysRevLett.87.160601},
   DOI={10.1103/physrevlett.87.160601},
   number={16},
   journal={Physical Review Letters},
   publisher={American Physical Society (APS)},
   author={Prokof’ev, Nikolay and Svistunov, Boris},
   year={2001},
   month=Sept }

@article{f4sj-rwvj,
  title = {Emergence of a boundary-sensitive phase in hyperbolic Ising models},
  author = {Wang, Xingzhi and Nussinov, Zohar and Ortiz, Gerardo},
  journal = {Phys. Rev. B},
  volume = {112},
  issue = {21},
  pages = {214102},
  numpages = {18},
  year = {2025},
  month = {Dec},
  publisher = {American Physical Society},
  doi = {10.1103/f4sj-rwvj},
  url = {https://link.aps.org/doi/10.1103/f4sj-rwvj}
}

@article{SciPostPhysCodeb.34,
	title = {HYPERTILING — a high performance Python library for the generation and visualization of hyperbolic lattices},
	pages = {34},
	author = {Schrauth, Manuel and Thurn, Yanick and Goth, Florian and Portela, Jefferson and Herdt, Dietmar and Dusel, Felix},
	journal = {SciPost Phys. Codebases},
	year = {2024},
	publisher = {SciPost},
	doi = {10.21468/SciPostPhysCodeb.34},
	url = {https://scipost.org/10.21468/SciPostPhysCodeb.34}
}

@article{SciPostPhysCodeb.34-r1.3,
	title = {Codebase release 1.3 for HYPERTILING},
	pages = {34-r1.3},
	author = {Schrauth, Manuel and Thurn, Yanick and Goth, Florian and Portela, Jefferson and Herdt, Dietmar and Dusel, Felix},
	journal = {SciPost Phys. Codebases},
	year = {2024},
	publisher = {SciPost},
	doi = {10.21468/SciPostPhysCodeb.34-r1.3},
	url = {https://scipost.org/10.21468/SciPostPhysCodeb.34-r1.3}
}

@article{Krcmar_2008,
   title={Ising model on a hyperbolic lattice studied by the corner transfer matrix renormalization group method},
   volume={41},
   ISSN={1751-8121},
   url={http://dx.doi.org/10.1088/1751-8113/41/12/125001},
   DOI={10.1088/1751-8113/41/12/125001},
   number={12},
   journal={Journal of Physics A: Mathematical and Theoretical},
   publisher={IOP Publishing},
   author={Krcmar, R and Gendiar, A and Ueda, K and Nishino, T},
   year={2008},
   month=Mar, pages={125001} }

@article{PhysRevD.103.094507,
  title = {Lattice setup for quantum field theory in ${\rm{AdS}}_{2}$},
  author = {Brower, Richard C. and Cogburn, Cameron V. and Fitzpatrick, A. Liam and Howarth, Dean and Tan, Chung-I},
  journal = {Phys. Rev. D},
  volume = {103},
  issue = {9},
  pages = {094507},
  numpages = {18},
  year = {2021},
  month = {May},
  publisher = {American Physical Society},
  doi = {10.1103/PhysRevD.103.094507},
  url = {https://link.aps.org/doi/10.1103/PhysRevD.103.094507}
}

@article{PhysRevD.106.054506,
  title = {Holography for Ising spins on the hyperbolic plane},
  author = {Asaduzzaman, Muhammad and Catterall, Simon and Hubisz, Jay and Nelson, Roice and Unmuth-Yockey, Judah},
  journal = {Phys. Rev. D},
  volume = {106},
  issue = {5},
  pages = {054506},
  numpages = {8},
  year = {2022},
  month = {Sep},
  publisher = {American Physical Society},
  doi = {10.1103/PhysRevD.106.054506},
  url = {https://link.aps.org/doi/10.1103/PhysRevD.106.054506}
}

@misc{morningstar2007monte,
      title={The Monte Carlo method in quantum field theory}, 
      author={Colin Morningstar},
      year={2007},
      eprint={hep-lat/0702020},
      archivePrefix={arXiv},
      primaryClass={hep-lat}
}

@article{Shima_2006,
   title={The dynamic exponent of the Ising model on negatively curved surfaces},
   volume={2006},
   ISSN={1742-5468},
   url={http://dx.doi.org/10.1088/1742-5468/2006/08/P08017},
   DOI={10.1088/1742-5468/2006/08/p08017},
   number={08},
   journal={Journal of Statistical Mechanics: Theory and Experiment},
   publisher={IOP Publishing},
   author={Shima, Hiroyuki and Sakaniwa, Yasunori},
   year={2006},
   month=Aug, pages={P08017–P08017} 
}

@article{PhysRevLett.99.110601,
  title = {Dynamic Critical Behavior of the Worm Algorithm for the Ising Model},
  author = {Deng, Youjin and Garoni, Timothy M. and Sokal, Alan D.},
  journal = {Phys. Rev. Lett.},
  volume = {99},
  issue = {11},
  pages = {110601},
  numpages = {4},
  year = {2007},
  month = {Sep},
  publisher = {American Physical Society},
  doi = {10.1103/PhysRevLett.99.110601},
  url = {https://link.aps.org/doi/10.1103/PhysRevLett.99.110601}
}

@article{PhysRevLett.125.053901,
  title = {Topological Hyperbolic Lattices},
  author = {Yu, Sunkyu and Piao, Xianji and Park, Namkyoo},
  journal = {Phys. Rev. Lett.},
  volume = {125},
  issue = {5},
  pages = {053901},
  numpages = {7},
  year = {2020},
  month = {Jul},
  publisher = {American Physical Society},
  doi = {10.1103/PhysRevLett.125.053901},
  url = {https://link.aps.org/doi/10.1103/PhysRevLett.125.053901}
}

@article{Wolff2009,
   title={Simulating the all-order strong coupling expansion I: Ising model demo},
   volume={810},
   ISSN={0550-3213},
   url={http://dx.doi.org/10.1016/j.nuclphysb.2008.09.033},
   DOI={10.1016/j.nuclphysb.2008.09.033},
   number={3},
   journal={Nuclear Physics B},
   publisher={Elsevier BV},
   author={Wolff, Ulli},
   year={2009},
   month=Apr, pages={491–502} }

@article{WOLFF2004143,
title = {Monte Carlo errors with less errors},
journal = {Computer Physics Communications},
volume = {156},
number = {2},
pages = {143-153},
year = {2004},
issn = {0010-4655},
doi = {https://doi.org/10.1016/S0010-4655(03)00467-3},
url = {https://www.sciencedirect.com/science/article/pii/S0010465503004673},
author = {Ulli Wolff}
}

@misc{Hesse2024pyuwerr,
  author  = {Hesse, Dirk},
  title   = {py-uwerr: Python implementation of Monte Carlo error analysis `a la Wolff},
  year    = {2024},
  url     = {https://github.com/dhesse/py-uwerr}
}

@article{Kulkarni2022,
doi = {10.1088/1742-5468/ac4c3e},
url = {https://doi.org/10.1088/1742-5468/ac4c3e},
year = {2022},
month = {feb},
publisher = {IOP Publishing and SISSA},
volume = {2022},
number = {2},
pages = {023209},
author = {Kulkarni, Suman and Dhar, Deepak},
title = {Finite size scaling functions of the phase transition in the ferromagnetic Ising model on random regular graphs},
journal = {Journal of Statistical Mechanics: Theory and Experiment}
}

@misc{dey2026boundaryconditiondependentuniversality,
      title={Boundary Condition dependent Universality Classes on a Hyperbolic Lattice}, 
      author={Pallabi Dey and Debasish Banerjee and Arnab Kundu and Ritam Sinha},
      year={2026},
      eprint={2609.37325},
      archivePrefix={arXiv},
      primaryClass={hep-lat},
      url={https://arxiv.org/abs/2609.37325}, 
}

@book{Cardy1996, 
  place={Cambridge}, 
  series={Cambridge Lecture Notes in Physics}, 
  title={Scaling and Renormalization in Statistical Physics}, 
  publisher={Cambridge University Press}, 
  author={Cardy, John}, 
  year={1996}, 
  collection={Cambridge Lecture Notes in Physics}
  }

@article{PhysRevLett.28.1516,
  title = {Scaling Theory for Finite-Size Effects in the Critical Region},
  author = {Fisher, Michael E. and Barber, Michael N.},
  journal = {Phys. Rev. Lett.},
  volume = {28},
  issue = {23},
  pages = {1516--1519},
  numpages = {0},
  year = {1972},
  month = {Jun},
  publisher = {American Physical Society},
  doi = {10.1103/PhysRevLett.28.1516},
  url = {https://link.aps.org/doi/10.1103/PhysRevLett.28.1516}
}

@article{Hasenbusch_2010,
   title={Finite size scaling study of lattice models in the three-dimensional Ising universality class},
   volume={82},
   ISSN={1550-235X},
   url={http://dx.doi.org/10.1103/PhysRevB.82.174433},
   DOI={10.1103/physrevb.82.174433},
   number={17},
   journal={Physical Review B},
   publisher={American Physical Society (APS)},
   author={Hasenbusch, Martin},
   year={2010},
   month=Nov }

@misc{bashmakov2025superconductivityhyperbolicspacesregular,
      title={Superconductivity in hyperbolic spaces: Regular hyperbolic lattices and Ginzburg-Landau theory}, 
      author={Vladimir Bashmakov and Askar Iliasov and Tomáš Bzdušek and Andrey A. Bagrov},
      year={2025},
      eprint={2509.09330},
      archivePrefix={arXiv},
      primaryClass={cond-mat.supr-con},
      url={https://arxiv.org/abs/2509.09330}, 
}

@article{Koll_r_2019,
   title={Hyperbolic lattices in circuit quantum electrodynamics},
   volume={571},
   ISSN={1476-4687},
   url={http://dx.doi.org/10.1038/s41586-019-1348-3},
   DOI={10.1038/s41586-019-1348-3},
   number={7763},
   journal={Nature},
   publisher={Springer Science and Business Media LLC},
   author={Kollár, Alicia J. and Fitzpatrick, Mattias and Houck, Andrew A.},
   year={2019},
   month=July, pages={45–50} }

@article{PhysRevD.105.114503,
  title = {Hyperbolic lattice for scalar field theory in ${\mathrm{AdS}}_{3}$},
  author = {Brower, Richard C. and Cogburn, Cameron V. and Owen, Evan},
  journal = {Phys. Rev. D},
  volume = {105},
  issue = {11},
  pages = {114503},
  numpages = {11},
  year = {2022},
  month = {Jun},
  publisher = {American Physical Society},
  doi = {10.1103/PhysRevD.105.114503},
  url = {https://link.aps.org/doi/10.1103/PhysRevD.105.114503}
}

@article{PhysRevD.102.034511,
  title = {Holography on tessellations of hyperbolic space},
  author = {Asaduzzaman, Muhammad and Catterall, Simon and Hubisz, Jay and Nelson, Roice and Unmuth-Yockey, Judah},
  journal = {Phys. Rev. D},
  volume = {102},
  issue = {3},
  pages = {034511},
  numpages = {9},
  year = {2020},
  month = {Aug},
  publisher = {American Physical Society},
  doi = {10.1103/PhysRevD.102.034511},
  url = {https://link.aps.org/doi/10.1103/PhysRevD.102.034511}
}

@article{Eggarter1974,
  title = {Cayley trees, the Ising problem, and the thermodynamic limit},
  author = {Eggarter, T. P.},
  journal = {Phys. Rev. B},
  volume = {9},
  issue = {7},
  pages = {2989--2992},
  numpages = {0},
  year = {1974},
  month = {Apr},
  publisher = {American Physical Society},
  doi = {10.1103/PhysRevB.9.2989},
  url = {https://link.aps.org/doi/10.1103/PhysRevB.9.2989}
}

@article{PhysRevLett.134.256604,
  title = {Chiral Gapless Spin Liquid in Hyperbolic Space},
  author = {Dusel, Felix and Hofmann, Tobias and Maity, Atanu and Mosseri, R{\'e}my and Vidal, Julien and Iqbal, Yasir and Greiter, Martin and Thomale, Ronny},
  journal = {Phys. Rev. Lett.},
  volume = {134},
  issue = {25},
  pages = {256604},
  numpages = {7},
  year = {2025},
  month = {Jun},
  publisher = {American Physical Society},
  doi = {10.1103/PhysRevLett.134.256604},
  url = {https://link.aps.org/doi/10.1103/PhysRevLett.134.256604}
}

@article{s25y-s4fj,
  title = {Hyperbolic Spin Liquids},
  author = {Lenggenhager, Patrick M. and Dey, Santanu and Bzdu{\ifmmode \check{s}\else \v{s}\fi{}}ek, Tom{\'a}{\ifmmode \check{s}\else \v{s}\fi{}} and Maciejko, Joseph},
  journal = {Phys. Rev. Lett.},
  volume = {135},
  issue = {7},
  pages = {076604},
  numpages = {10},
  year = {2025},
  month = {Aug},
  publisher = {American Physical Society},
  doi = {10.1103/s25y-s4fj},
  url = {https://link.aps.org/doi/10.1103/s25y-s4fj}
}

@book{privman1990finite,
  title={Finite Size Scaling and Numerical Simulation of Statistical Systems},
  author={Privman, V.},
  isbn={9789813208766},
  year={1990},
  publisher={World Scientific Publishing Company}
}

@article{PRIVMAN1991241,
title = {Finite-size scaling: new results},
journal = {Physica A: Statistical Mechanics and its Applications},
volume = {177},
number = {1},
pages = {241-246},
year = {1991},
issn = {0378-4371},
doi = {https://doi.org/10.1016/0378-4371(91)90159-A},
url = {https://www.sciencedirect.com/science/article/pii/037843719190159A},
author = {Vladimir Privman}
}

@article{PhysRevB.74.144506,
  title = {Theoretical estimates of the critical exponents of the superfluid transition in $^{4}\mathrm{He}$ by lattice methods},
  author = {Campostrini, Massimo and Hasenbusch, Martin and Pelissetto, Andrea and Vicari, Ettore},
  journal = {Phys. Rev. B},
  volume = {74},
  issue = {14},
  pages = {144506},
  numpages = {18},
  year = {2006},
  month = {Oct},
  publisher = {American Physical Society},
  doi = {10.1103/PhysRevB.74.144506},
  url = {https://link.aps.org/doi/10.1103/PhysRevB.74.144506}
}

@article{PhysRevE.72.016128,
  title = {Surface and bulk transitions in three-dimensional $\mathrm{O}(n)$ models},
  author = {Deng, Youjin and Bl{\"o}te, Henk W. J. and Nightingale, M. P.},
  journal = {Phys. Rev. E},
  volume = {72},
  issue = {1},
  pages = {016128},
  numpages = {11},
  year = {2005},
  month = {Jul},
  publisher = {American Physical Society},
  doi = {10.1103/PhysRevE.72.016128},
  url = {https://link.aps.org/doi/10.1103/PhysRevE.72.016128}
}

@article{Boettcher_2020,
   title={Quantum simulation of hyperbolic space with circuit quantum electrodynamics: From graphs to geometry},
   volume={102},
   ISSN={2469-9934},
   url={http://dx.doi.org/10.1103/PhysRevA.102.032208},
   DOI={10.1103/physreva.102.032208},
   number={3},
   journal={Physical Review A},
   publisher={American Physical Society (APS)},
   author={Boettcher, Igor and Bienias, Przemyslaw and Belyansky, Ron and Kollár, Alicia J. and Gorshkov, Alexey V.},
   year={2020},
   month=Sept 
}

@article{PhysRevLett.128.013601,
  title = {Circuit Quantum Electrodynamics in Hyperbolic Space: From Photon Bound States to Frustrated Spin Models},
  author = {Bienias, Przemyslaw and Boettcher, Igor and Belyansky, Ron and Koll{\'a}r, Alicia J. and Gorshkov, Alexey V.},
  journal = {Phys. Rev. Lett.},
  volume = {128},
  issue = {1},
  pages = {013601},
  numpages = {7},
  year = {2022},
  month = {Jan},
  publisher = {American Physical Society},
  doi = {10.1103/PhysRevLett.128.013601},
  url = {https://link.aps.org/doi/10.1103/PhysRevLett.128.013601}
}

@article{Saa_2026,
   title={Higher-dimensional Euclidean and non-Euclidean structures in planar circuit quantum electrodynamics},
   volume={113},
   ISSN={2469-9969},
   url={http://dx.doi.org/10.1103/z2zk-m4c3},
   DOI={10.1103/z2zk-m4c3},
   number={3},
   journal={Physical Review B},
   publisher={American Physical Society (APS)},
   author={Saa, Alberto and Miranda, Eduardo and Rouxinol, Francisco},
   year={2026},
   month=Jan }

@article{Sun_2024,
   title={Topological linear response of hyperbolic Chern insulators},
   volume={17},
   ISSN={2542-4653},
   url={http://dx.doi.org/10.21468/SciPostPhys.17.5.124},
   DOI={10.21468/scipostphys.17.5.124},
   number={5},
   journal={SciPost Physics},
   publisher={Stichting SciPost},
   author={Sun, Canon and Chen, Anffany and Bzdusek, Tomas and Maciejko, Joseph},
   year={2024},
   month=Nov }

@article{Shankar_2024,
   title={Hyperbolic Lattices and Two-Dimensional Yang-Mills Theory},
   volume={133},
   ISSN={1079-7114},
   url={http://dx.doi.org/10.1103/PhysRevLett.133.146601},
   DOI={10.1103/physrevlett.133.146601},
   number={14},
   journal={Physical Review Letters},
   publisher={American Physical Society (APS)},
   author={Shankar, G. and Maciejko, Joseph},
   year={2024},
   month=Sept }

@article{PhysRevLett.131.226401,
  title = {Non-Abelian Hyperbolic Band Theory from Supercells},
  author = {Lenggenhager, Patrick M. and Maciejko, Joseph and Bzdu{\ifmmode \check{s}\else \v{s}\fi{}}ek, Tom{\'a}{\ifmmode \check{s}\else \v{s}\fi{}}},
  journal = {Phys. Rev. Lett.},
  volume = {131},
  issue = {22},
  pages = {226401},
  numpages = {7},
  year = {2023},
  month = {Dec},
  publisher = {American Physical Society},
  doi = {10.1103/PhysRevLett.131.226401},
  url = {https://link.aps.org/doi/10.1103/PhysRevLett.131.226401}
}

@article{Maciejko_2022,
   title={Automorphic Bloch theorems for hyperbolic lattices},
   volume={119},
   ISSN={1091-6490},
   url={http://dx.doi.org/10.1073/pnas.2116869119},
   DOI={10.1073/pnas.2116869119},
   number={9},
   journal={Proceedings of the National Academy of Sciences},
   publisher={National Academy of Sciences},
   author={Maciejko, Joseph and Rayan, Steven},
   year={2022},
   month=Feb }

@article{Boettcher_2022,
   title={Crystallography of hyperbolic lattices},
   volume={105},
   ISSN={2469-9969},
   url={http://dx.doi.org/10.1103/PhysRevB.105.125118},
   DOI={10.1103/physrevb.105.125118},
   number={12},
   journal={Physical Review B},
   publisher={American Physical Society (APS)},
   author={Boettcher, Igor and Gorshkov, Alexey V. and Kollár, Alicia J. and Maciejko, Joseph and Rayan, Steven and Thomale, Ronny},
   year={2022},
   month=Mar }

@article{Maldacena_1999,
   title={The Large-N Limit of Superconformal Field Theories and Supergravity},
   volume={38},
   ISSN={1572-9575},
   url={http://dx.doi.org/10.1023/A:1026654312961},
   DOI={10.1023/a:1026654312961},
   number={4},
   journal={International Journal of Theoretical Physics},
   publisher={Springer Science and Business Media LLC},
   author={Maldacena, Juan},
   year={1999},
   month=Apr, pages={1113–1133} }

@misc{witten1998antisitterspaceholography,
      title={Anti De Sitter Space And Holography}, 
      author={Edward Witten},
      year={1998},
      eprint={hep-th/9802150},
      archivePrefix={arXiv},
      primaryClass={hep-th},
      url={https://arxiv.org/abs/hep-th/9802150}, 
}

@article{Basteiro_2022,
   title={Towards explicit discrete holography:  Aperiodic spin chains from hyperbolic tilings},
   volume={13},
   ISSN={2542-4653},
   url={http://dx.doi.org/10.21468/SciPostPhys.13.5.103},
   DOI={10.21468/scipostphys.13.5.103},
   number={5},
   journal={SciPost Physics},
   publisher={Stichting SciPost},
   author={Basteiro, Pablo and Di Giulio, Giuseppe and Erdmenger, Johanna and Karl, Jonathan and Meyer, Rene and Xian, Zhuo-Yu},
   year={2022},
   month=Nov }

@article{Boyle_2025,
   title={Holographic foliations: Self-similar quasicrystals from hyperbolic honeycombs},
   volume={111},
   ISSN={2470-0029},
   url={http://dx.doi.org/10.1103/PhysRevD.111.046001},
   DOI={10.1103/physrevd.111.046001},
   number={4},
   journal={Physical Review D},
   publisher={American Physical Society (APS)},
   author={Boyle, Latham and Kulp, Justin},
   year={2025},
   month=Feb }

@article{Jahn_2020,
   title={Central charges of aperiodic holographic tensor-network models},
   volume={102},
   ISSN={2469-9934},
   url={http://dx.doi.org/10.1103/PhysRevA.102.042407},
   DOI={10.1103/physreva.102.042407},
   number={4},
   journal={Physical Review A},
   publisher={American Physical Society (APS)},
   author={Jahn, Alexander and Zimborás, Zoltán and Eisert, Jens},
   year={2020},
   month=Oct }

@article{Gesteau_2022,
   title={Holographic tensor networks from hyperbolic buildings},
   volume={2022},
   ISSN={1029-8479},
   url={http://dx.doi.org/10.1007/JHEP10(2022)169},
   DOI={10.1007/jhep10(2022)169},
   number={10},
   journal={Journal of High Energy Physics},
   publisher={Springer Science and Business Media LLC},
   author={Gesteau, Elliott and Marcolli, Matilde and Parikh, Sarthak},
   year={2022},
   month=Oct }

@article{Jahn_2022,
   title={Tensor network models of AdS/qCFT},
   volume={6},
   ISSN={2521-327X},
   url={http://dx.doi.org/10.22331/q-2022-02-03-643},
   DOI={10.22331/q-2022-02-03-643},
   journal={Quantum},
   publisher={Verein zur Forderung des Open Access Publizierens in den Quantenwissenschaften},
   author={Jahn, Alexander and Zimborás, Zoltán and Eisert, Jens},
   year={2022},
   month=Feb, pages={643} }

@article{Breuckmann_2016,
   title={Constructions and Noise Threshold of Hyperbolic Surface Codes},
   volume={62},
   ISSN={1557-9654},
   url={http://dx.doi.org/10.1109/TIT.2016.2555700},
   DOI={10.1109/tit.2016.2555700},
   number={6},
   journal={IEEE Transactions on Information Theory},
   publisher={Institute of Electrical and Electronics Engineers (IEEE)},
   author={Breuckmann, Nikolas P. and Terhal, Barbara M.},
   year={2016},
   month=June, pages={3731–3744} }

@misc{taylor2023holographycellulationserrorcorrecting,
      title={Holography, cellulations and error correcting codes}, 
      author={Marika Taylor and Charles Woodward},
      year={2023},
      eprint={2112.12468},
      archivePrefix={arXiv},
      primaryClass={hep-th},
      url={https://arxiv.org/abs/2112.12468}, 
}

@misc{jaroszewicz2026temperaturescalesisingmodel,
      title={Two temperature scales in the Ising model on the $\{5,4\}$ hyperbolic lattice with free boundaries: susceptibility peak and boundary-induced order}, 
      author={Sebastian Jaroszewicz and Nahuel Mendez and Maria P. Beccar-Varela and Maria C. Mariani},
      year={2026},
      eprint={2609.29323},
      archivePrefix={arXiv},
      primaryClass={cond-mat.stat-mech},
      url={https://arxiv.org/abs/2609.29323}, 
}

@misc{samlodia2026quantumisingmodel21dimensional,
      title={Quantum Ising Model on $(2+1)-$Dimensional Anti$-$de Sitter Space using Tensor Networks}, 
      author={Abhishek Samlodia and Simon Catterall and Alexander F. Kemper and Yannick Meurice and Goksu Can Toga},
      year={2026},
      eprint={2512.20838},
      archivePrefix={arXiv},
      primaryClass={hep-lat},
      url={https://arxiv.org/abs/2512.20838}, 
}

@article{Barthold2002,
   title={Salem Numbers and Growth Series of Some Hyperbolic Graphs},
   volume={90},
   ISSN={1572-9168},
   url={http://dx.doi.org/10.1023/A:1014902918849},
   DOI={10.1023/a:1014902918849},
   number={1},
   journal={Geometriae Dedicata},
   publisher={Springer Science and Business Media LLC},
   author={Barthold, Laurent and Ceccherini-Silberstein, Tullio G.},
   year={2002},
   month=Mar, pages={107–114} }

@ARTICLE{1981ZPhyB..43..119B,
       author = {{Binder}, K.},
        title = "{Finite size scaling analysis of ising model block distribution functions}",
      journal = {Zeitschrift fur Physik B Condensed Matter},
         year = 1981,
        month = jun,
       volume = {43},
       number = {2},
        pages = {119-140},
          doi = {10.1007/BF01293604},
       adsurl = {https://ui.adsabs.harvard.edu/abs/1981ZPhyB..43..119B}
}

@article{10.1063/1.3518900,
    author = {Sandvik, Anders W.},
    title = {Computational Studies of Quantum Spin Systems},
    journal = {AIP Conference Proceedings},
    volume = {1297},
    number = {1},
    pages = {135-338},
    year = {2010},
    month = {11},
    issn = {0094-243X},
    doi = {10.1063/1.3518900},
    url = {https://doi.org/10.1063/1.3518900}
}

@article{Banerjee_2021,
   title={Recent progress on cluster and meron algorithms for strongly correlated systems},
   volume={95},
   ISSN={0974-9845},
   url={http://dx.doi.org/10.1007/s12648-021-02155-5},
   DOI={10.1007/s12648-021-02155-5},
   number={8},
   journal={Indian Journal of Physics},
   publisher={Springer Science and Business Media LLC},
   author={Banerjee, Debasish},
   year={2021},
   month=June, pages={1669–1680} }

@article{Hasenbusch_1999,
   title={A Monte Carlo study of leading order scaling corrections of $\phi^4$ theory on a three-dimensional lattice},
   volume={32},
   ISSN={1361-6447},
   url={http://dx.doi.org/10.1088/0305-4470/32/26/304},
   DOI={10.1088/0305-4470/32/26/304},
   number={26},
   journal={Journal of Physics A: Mathematical and General},
   publisher={IOP Publishing},
   author={Hasenbusch, M},
   year={1999},
   month=Jan, pages={4851–4865} }

@article{PhysRevB.44.5081,
  title = {Critical behavior of the three-dimensional Ising model: A high-resolution Monte Carlo study},
  author = {Ferrenberg, Alan M. and Landau, D. P.},
  journal = {Phys. Rev. B},
  volume = {44},
  issue = {10},
  pages = {5081--5091},
  numpages = {0},
  year = {1991},
  month = {Sep},
  publisher = {American Physical Society},
  doi = {10.1103/PhysRevB.44.5081},
  url = {https://link.aps.org/doi/10.1103/PhysRevB.44.5081}
}

@article{RevModPhys.49.435,
  title = {Theory of dynamic critical phenomena},
  author = {Hohenberg, P. C. and Halperin, B. I.},
  journal = {Rev. Mod. Phys.},
  volume = {49},
  issue = {3},
  pages = {435--479},
  numpages = {0},
  year = {1977},
  month = {Jul},
  publisher = {American Physical Society},
  doi = {10.1103/RevModPhys.49.435},
  url = {https://link.aps.org/doi/10.1103/RevModPhys.49.435}
}

@book{texbook,
    author = {Montvay I and Münster G} ,
    title = {Quantum Fields on a Lattice},
    publisher = {Cambridge Monographs on Mathematical Physics},
    chapter = {1,2,7},
    year = {1994}
}

@book{montecarlo_stat,
    author = {M. E. J. Newman \& G. T. Barkema},
    title = {Monte Carlo Methods in Statistical Physics},
    publisher = {Oxford University Press},
    year = {1999} 
}

@Inbook{Janke2008,
author="Janke, Wolfhard",
editor="Fehske, H.
and Schneider, R.
and Wei{\ss}e, A.",
title="Monte Carlo Methods in Classical Statistical Physics",
bookTitle="Computational Many-Particle Physics",
year="2008",
publisher="Springer Berlin Heidelberg",
address="Berlin, Heidelberg",
pages="79--140",
isbn="978-3-540-74686-7",
doi="10.1007/978-3-540-74686-7_4",
url="https://doi.org/10.1007/978-3-540-74686-7_4"
}

@article{Joswig_2023,
   title={pyerrors: A python framework for error analysis of Monte Carlo data},
   volume={288},
   ISSN={0010-4655},
   url={http://dx.doi.org/10.1016/j.cpc.2023.108750},
   DOI={10.1016/j.cpc.2023.108750},
   journal={Computer Physics Communications},
   publisher={Elsevier BV},
   author={Joswig, Fabian and Kuberski, Simon and Kuhlmann, Justus T. and Neuendorf, Jan},
   year={2023},
   month=July, pages={108750} }
\newpage
\appendix

\end{document}